\documentclass[10pt,journal,compsoc]{IEEEtran}

\usepackage[numbers,sort&compress]{natbib}
\usepackage[switch]{lineno}
\usepackage[table]{xcolor}
\usepackage{stfloats}
\usepackage{diagbox}
\usepackage{graphicx}
\usepackage{amsmath}
\usepackage{multicol}
\usepackage{amssymb}
\usepackage{multirow}
\usepackage{epstopdf}
\usepackage{ifpdf}
\usepackage{amsfonts}
\usepackage{amsthm}
\usepackage{mathrsfs}
\usepackage{makecell}
\usepackage{color}
\usepackage{bm}

\usepackage[caption=false]{subfig}
\usepackage[linesnumbered,ruled,vlined]{algorithm2e}

\newcommand{\Square[1]}{{#1}}

\newtheorem{theorem}{Theorem}

\newtheorem{remark}{Remark}

\graphicspath{{figures/}}
\begin{document}
	
	\title{Modeling and Detecting Communities in \\ Node Attributed Networks}

	\author{
		{Ren Ren, Jinliang Shao, Adrian N. Bishop,~\IEEEmembership{Senior Member, IEEE},
			Wei Xing Zheng,~\IEEEmembership{Fellow, IEEE}}
		\thanks{This work was supported in part by the National Science Foundation of China under Grant U1830207, Grant 61772003, and Grant 61903066; in part by the Sichuan Science and Technology Program under Grant 2021YFH0042; in part by the Shenzhen Institute of Artificial Intelligence and Robotics for Society; and in part by the NSW Cyber Security Network in Australia under Grant P00025091.}
		\thanks{Ren Ren is with School of Automation Engineering, University of Electronic Science and Technology of China, 611731, China (e-mail: r.ren.scholar@gmail.com).}
		\thanks{Jinliang Shao is with School of Automation Engineering, University of Electronic Science and Technology of China, 611731, China, and also with the Research Center on Crowd Spectrum Intelligence, Shenzhen Institute of Artificial Intelligence and Robotics for Society, Shenzhen 518054, China (e-mail:jinliangshao@uestc.edu.cn).}
		\thanks{Adrian Bishop is with University of Technology Sydney (UTS), Australia, and he is also with Data61 (CSIRO) Canberra Research Lab, Australia (e-mail: adrian.bishop@uts.edu.au).}
		\thanks{W. X. Zheng is with the School of Computer, Data and Mathematical Sciences, Western Sydney University, Sydney, NSW 2751, Australia (e-mail: w.zheng@westernsydney.edu.au).}
	}

\IEEEtitleabstractindextext
{
\vspace*{-1mm}
\begin{abstract}
As a fundamental structure in  real-world networks,  in addition to graph topology, communities can also be  reflected by    abundant node attributes.  In {attributed} community detection,  probabilistic generative models (PGMs) have become the mainstream  method due to their principled characterization and competitive performances.
Here, we propose a novel PGM without imposing any distributional assumptions on attributes, which is superior to the existing PGMs that require attributes to be categorical or Gaussian distributed.
Based on the block model  {of graph structure}, our model incorporates the attribute by describing its effect on node popularity. To characterize the effect quantitatively, we analyze the community detectability  for our model and then establish the requirements of the node popularity term.  {This leads to a new scheme for the crucial model selection  problem in choosing and solving attributed community detection models}.  
With the model determined, an efficient algorithm is developed to estimate the parameters and to infer the communities. 
The proposed method is validated from two aspects. First, the effectiveness of our algorithm  is theoretically guaranteed by  the detectability condition. {Second, extensive experiments  indicate that our method not only outperforms the competing approaches on the employed datasets, but also shows  better applicability to networks with various node attributes.}
\end{abstract} 
}

\maketitle

\section{Introduction}
Many real-world complex systems naturally form multiple groups of individuals  with close relationships or strong similarity, instances of which include social  circles of online users,  functional modules constructed by interacting proteins, etc \cite{yang2015defining,Fortunato2016}. Abstracting the system as a network with nodes and edges, the concept ``community'' was proposed to depict the   assortative structural groups/modules where the nodes have more links to others in the same group  than the rest of the network \cite{Girvan2002a}, whose detection has become a fundamental tool in network analysis.  However, the links in real-world networks are often sparse and noisy \cite{Newman2018a}, which may  depress the performance of community detection \cite{Hric2014} or even make the communities essentially undetectable \cite{Decelle2011a,Moore2017}. 

Fortunately,  in addition to the structural information, most real-world networks contain abundant node attributes,  e.g., the co-purchasing network annotated by product categories \cite{yang2015defining,Hric2014},  which can not only  reflect the similarity between nodes, but may also even directly indicate the community memberships. Nevertheless it is notable that using the attribute only is rarely adequate to reveal the network modules. In fact, the labeled categories are often too coarse to classify  the products in Amazon \cite{Fortunato2016,Hric2014}. 

In order to take full advantages of the useful information in real-world networks, great effort has been devoted to the fusion of graph structure and node attribute data in network analysis, raising the research topic of attributed community detection \cite{chunaev2020community}. Among a variety of data fusion approaches, the  probabilistic generative model (PGM)-based methods {have shown very competitive and robust performances} 
\cite{myth} and have become the mainstream  \cite{Yang2013,Hric2016b}. In the language of probability, PGMs clearly describe the dependence of networks on different factors such as latent groups and node degrees in a principled way \cite{myth}, and thus can be used to quantify the correlation between attributes and communities \cite{Newman2016a}, to prove the performance of algorithms \cite{Peel2016,Zhang2014shui}, to reveal the functions of modules \cite{CHANG2019252,joint}, {and to make direct comparisons between models \cite{myth}. }

One of the significant advantages of the PGM is that it allows principled analysis on  the condition  of communities' being detected, i.e.,  the so-called detectability of communities, {which plays a central role in the statistical descriptions of the significance of community structure} \cite{Decelle2011a,Moore2017, myth}.  For node attributed networks, the pioneering work \cite{Zhang2014shui} showed in general that a fraction of nodes with known memberships can improve the detectability, using the topology-based algorithm in \cite{Decelle2011a}.   And the detectability analysis for a specific attribute-aware model     was {empirically} performed in \cite{Newman2016a}, which also validated the effectiveness of the proposed method thereof.

Based on the Stochastic Block Model (SBM), which generates network edges according to the latent block structure and the group membership of nodes \cite{Holland1983}, two schemes are usually adopted in the existing PGMs to integrate node attributes.
One scheme models the generative process of both edges and attribute vectors \cite{Yang2013,Xu2012,CHANG2019252,Zanghi2010,ADD2}, which usually requires the distribution of attributes to be specified. For example, it is assumed in some models that categorical attributes follow a multinomial or binary distribution \cite{Yang2013,Xu2012,CHANG2019252,ADD2} and continuous ones obey a multivariate Gaussian distribution \cite{Zanghi2010}.
The other scheme only focuses on the generation of edges and  {the data fusion} is manifested by the dependence of link possibilities on attributes \cite{Newman2016a,Hric2016b,joint}, where the attributes are seen as given parameters. By this means, these works incorporate categorical or univariate continuous attributes into analysis, while multidimensional real-valued ones have not been tackled.

In fact, node attributes in real-life networks  often contain multidimensional and continuous values \cite{chunaev2020community}, {whose typical instances include word embeddings in   citation graphs \cite{mag} and locations in transport networks \cite{Simini2012a}. In this case, PGMs that can only handle categorical attributes may not be  adequately cooperated  with existing data mining technologies such as topic modeling \cite{topic}.}
Despite that real-world data appeal to  PGMs  for various node attributes, the development of such models is still an open problem addressed by few papers, as pointed in \cite{Hric2016b}. Furthermore, for the design and inference of PGMs, an inherent issue is the principled choice of different models \cite{Newman2016a}. Currently, such choice is usually conducted according to {prior knowledge about the generating procedure of attributes} \cite{Yang2013,Xu2012,CHANG2019252,ms} or model selection criteria \cite{Newman2016a,Hric2016b,ms}.  But for diverse real-valued or mixed attributes, the challenge lies  in that, it is hardly possible to specify a universal and reasonable prior distribution or generating process. Consequently, the widely used nonparametric Bayesian technologies and   model selection criteria  \cite{Yan2014,Peixoto2014,ms} are also hard to be applied. 

In this paper, we propose a novel PGM to model  communities with the fusion of edges and node attributes, and then it can be routinely  applied to community detection via model inference. 
For the generality of our model,  no distributional assumption is imposed on  attributes and the challenging model selection problem is instead addressed through a principled algorithmic analysis. 
In detail, we focus on the generation of edges depending on the blocks and the distances between node attributes, so that  communities are  highly correlated to both attributes and graph structure. Based on SBM, the primary issue is to choose a model that effectively characterizes the dependence  of  edges on node  attributes, or from another viewpoint, the effect of attributes on linking possibilities. To this end, we investigate the detectability condition of communities in attributed networks for the proposed model. The detectability analysis provides a quantitative description on the effect of node attributes,  thus leading to a novel model selection scheme.

The main    contributions in this paper can be summarized  as: 1)  {We propose a new Bayesian generative model  for community detection that can incorporate either categorical or real-valued node attributes}, {so that more information in real-life networks such as words can be fused.}  2) We  analyze the community detectability   for the proposed model {and compare it with that of the topology-based counterpart,  thus clarifying the effect of   attributes on community detection.} 3) We present a novel model selection scheme and  develop efficient algorithms to estimate the parameters and to infer the communities.
Finally, we perform numerical  experiments  on artificial networks to verify the detectability analysis, and conduct experiments on extensive real-world datasets to demonstrate the superior performance of our algorithm.
 
\section{Related Work}
{Community detection has been a hot topic in network analysis and the methods proposed for this task are really numerous \cite{Fortunato2016}. We here only introduce the most related literature with our work and  focus on models and algorithms.  Readers can refer to \cite{chunaev2020community,Fortunato2016} for a comprehensive survey.}

{With the development of this field, SBM has played a central role in algorithm design and  analysis \cite{Peixotonp}. It has also been  shown that SBM-based approaches are equivalent to modularity optimization \cite{equ} and spectral methods \cite{Krzakala2013} in some cases. SBM assumes that the modular network  can be divided into blocks according to the hidden communities and the linking possibility of nodes is determined by the block structure. To describe the heterogeneous vertex degrees,  the Degree Corrected SBM (DCSBM) \cite{Karrer2011} further adds the node degree term into the model. Moreover, SBM has been  extended to various cases including networks with hierarchical communities and multiplex edges, etc \cite{Peixotonp}.  To solve and compare different models, model selection criteria such as Minimum Description Length (MDL) \cite{Peixoto2014} and Factorized Information Criteria  \cite{ms} are widely applied.}

{When it comes to attributed community extraction, PGM based methods depict the generation of edges based on SBM, while the modeling of attributes can be roughly classified into two kinds. One line of approaches, such as BAGC \cite{Xu2012}, BTLSC \cite{CHANG2019252} and CohsMix \cite{Zanghi2010}, generate the attributes conditioned on the blocks and indeed specify the distribution of attributes, which can be binary \cite{Xu2012}, multinomial \cite{CHANG2019252}, or Gaussian \cite{Zanghi2010}. Such scheme results in complicated hierarchical Bayesian models \cite{CHANG2019252,ADD2}, which are often solved by nonparametric Bayesian technologies \cite{ms}.}

{The other line of studies including SI  \cite{Newman2016a} and LSBM \cite{Hric2016b} are more relevant to our paper. In these works, attributes are not fused according to their generation process but instead are treated as known data or parameters that  determine edges jointly with degrees and blocks \cite{Newman2016a,Hric2016b,joint}. Take SI as an example. SI integrates a set of alignment parameters for each pair of community and   attribute and the resulting SI model is the product of DCSBM and alignment parameters. However, these models mainly focus on discrete node features, and as discussed in \cite{Newman2016a}, the incorporation of real-valued attributes still faces serious model selection problem.}

{Besides probabilistic models,  the relation between latent groups and edges/attributes can also be   explicitly described by non-negative matrix factorization (NMF) models  \cite{ascd, sci}. For example, SCI \cite{sci}  approximates both the graph adjacency and node feature matrix by linear combinations (or  particularly, inner product) of community memberships respectively to obtain a unified optimization objective for node clustering.  Further, the linear combinations can be  extended to predefined or neural-network-based nonlinear transforms so that the graph structure is embedded into a new space, leading to network embedding approaches \cite{nec,cde,emb}. However, for both NMF and embedding methods, the balance weights of different terms with respect to edges and attributes in the objective function are hard to decide for a unsupervised clustering task \cite{chunaev2020community}.}

Additionally,  node-augmented graph-based methods should be included for the completeness of literature review. These algorithms directly model the influence of attributes   by adding new nodes and edges to the original graph according to elaborate metrics and rules \cite{inc}, which contrasts to PGMs  that fit the given network data.

\section{The proposed model}
	\textit{Notations:} An undirected binary network  with $ n  $ annotated nodes and $ m $ edges can be denoted by $ G = (V,E, X)$, where $ V $ is the  node set,   $ E \subseteq V \times V $ is the edge set, and $ X = \{\mathbf{x}_i | \mathbf{x}_i \in \mathbb{R}^{d}, i \in V \}$ is the set of  $ d $ dimensional node attributes. 
	Let $ z_i \in [q] $ be the   membership of node $ i $, where $ [q] $ is the shorthand of  the set $ \{1,2,\ldots,q\}  $ and $ q $ is the number of communities in $ G $.  Besides, we further define $ \mathcal{C}_r  \triangleq \{ \mathbf{x}_\ell \mid \ell \in V, z_\ell = r \}$ to be the cluster composed of the attributes of the nodes in the community $ r $. We note that for clarity,   $ l $, $ i $ and $ j $ are used to index nodes, and $ r $, $ s $, $ u $, $ v $ to index communities throughout this paper.  Other notations will be explained in the context.
	A list of notations involved in this work is also summarized
	in Table 1 for convenience.
	
\begin{table}[t]
\renewcommand\arraystretch{1.1}
	\caption{ List of Notations}
	\vspace*{-2mm}
	\centering
	\begin{tabular}{l|l }
		\hline
Symbol & Description \\
\hline
$ G = (V,E, X) $              & $ G $: graph, $ V $: node set, $ E $: edge set, $ X $:  attributes \\
$ p_{ij} $, $ a_{ij} $ & $ p_{ij} $: link possibility, $ a_{ij} $: adjacency matrix entry \\
$ \mathbf{z} = (z_1, \ldots, z_n) $ & vector of community membership \\

\hline
$  \mathcal{C}_r  $ & attributes of nodes in community $ r $ \\
$ \bm{\zeta}_{r} $ & cluster center of $  \mathcal{C}_r $\\
$ \alpha_{ir} $ & normalized distance of  $ i $ and   $\bm{\zeta}_{r}$\\
$ \omega $, $ g $ & $ \omega $:  parameter for blocks, $ g $: node popularity \\
$ f_{ir} $  & abbreviation of $ f(\alpha_{ir}) $ where $ f $ is a function \\
$ \bm \beta$, $ \vartheta  $ & parameter set of   $ f(\cdot) $ and CRSBM, respectively \\
$ \Xi_{rs} $ & sum of $ g_{ij} $ for node pairs in block $ (r,s) $\\
$ n_s^r $ & sum of $ f(\alpha_{ir}) $ for every node  $ i $ in group $ s $\\
\hline
$ \psi $, $ h $ & $ \psi $: message or belief, $ h $: auxiliary external field \\
$  W_{sr}^{\ell i}  $ & coupling weight between groups $ s $ and $ r $ in BP \\
\hline
	\end{tabular}
	\vspace*{-2mm}
	\label{table0}
\end{table}

\subsection{Model Description}
{
	In general, the graph  topology of $ G $ can be generated by a family of model where each edge $(i,j)\in E $   is  independently generated via a Bernoulli distribution parameterized by a possibility $ p_{ij} $ \cite{Holland1983}. Then it follows the likelihood
	\begin{equation}\label{Eq1}
	P(G|\vartheta) = \prod_{i < j}p_{ij}^{a_{ij}}(1-p_{ij})^{1-a_{ij}},
	\end{equation}
	where   $ \vartheta $ is the parameter set of the  model, and $ a_{ij} = 1 $ if there is an edge between $ i $ and $ j $, otherwise $ 0 $.}

Based on the model family (\ref{Eq1}), the SBM  assumes that the network with $ q $ planted communities can be divided into $ q \times q $ blocks and the linking possibilities in the same block are equal, i.e., $ p_{ij} = \omega_{z_i,z_j} $ with $ \omega_{z_i,z_j} $ being the  edge density of the block $ (z_i,z_j) \in  [q]  \times  [q] $, which generates an Erd\"{o}s-R\'{e}nyi (ER) graph with Poissonian degree distribution.
 
To describe networks with arbitrary degree distributions, DCSBM  assumes $ p_{ij} = g_{ij} \omega_{z_i,z_j}\! =\! k_ik_j \omega_{z_i,z_j}$, with $ k_i $ being the degree of node $ i $  \cite{Karrer2011,Peixotonp}. 
Besides the term $ \omega $ that describes the block structure, DCSBM further characterizes the linking possibility $ p_{ij} $  by another term $ k_ik_j $ with respect to the individual property of each node of the endpoint pair. Indeed, the  degree $ k $ naturally reflects  the so-called \textit{popularity} of the node, that is, the tendency or likelihood of a node establishing connections with  other nodes  \cite{Faqeeh2018}.  From this viewpoint, the degree correction is in line with the intuition that a pair of agents are more likely to be linked if  they both have high popularity. This motivates us to model  $ p_{ij} $ using   available features of the node pair $ (i,j) $ in addition to the block term.  

A second inspiration comes from the  existing studies showing that the connections between nodes  are largely determined by their distances or  differences in some real-world networks. For instance, the flow volume between two places  decreases as their  geographical distance increases \cite{Simini2012a}. Considering this, a straightforward   extension of SBM  for node attributed networks is  
\begin{equation}\label{model0}
p_{ij} = g_{ij} \omega_{z_i,z_j}  \text{~with~} g_{ij} = f(\| \mathbf{x}_i- \mathbf{x}_j\|).
\end{equation}
By setting $ f $ as a real-valued function of the distance between   attributes, this model can tackle  categorical, real  and mixed-valued attributes.

However, the distances of every node pair are  usually sensitive to  noise and expensive to compute \cite{steinbach2005cluster}. To overcome these drawbacks, sparked by  DCSBM, we propose a novel model where $ g_{ij} $ is the product of the node-wise popularity of  $ i $ and $ j $.  Let   $ \bm{\zeta}_{r} $ denote the cluster representative   prototype (CRP) \cite{steinbach2005cluster} {or weighted cluster center} of  the cluster $ \mathcal{C}_r $ of node attributes, and let
\begin{equation}\label{alpha}
\alpha_{ir} =  {\Square[{\|\mathbf{x}_i-\bm{\zeta}_{r}\|}]}{\Big / }{\sum\nolimits_{r=1}^q\Square[{\|\mathbf{x}_i-\bm{\zeta}_{r}\|}]}
\end{equation}
denote the normalized distance between node $ i $ and cluster $  \mathcal{C}_r $. Our model can be then  written as $  p_{ij} = g_{ij} \omega_{z_i,z_j}  $ with  
\begin{equation}\label{gravity}
g_{ij} = f(\alpha_{i,z_j})\cdot f(\alpha_{j,z_i}),
\end{equation}
where the real-valued function  $ f $    describes node popularity. {By this means, we fuse both node attributes and graph topology into the generation of network communities, and $ f $ partly determines the relative weight of attributes in the model.} In Eq.~(\ref{gravity}), the  distances between $ O(qn) $ pairs of attributes and CRPs are used to replace those   between $ O(n^2) $ attribute pairs in \eqref{model0}, which describes that the linking possibility of a node pair is partly determined by the distance between one's attribute and the other's  cluster. 
Such strategy is in the spirit of the classical data clustering algorithm k-means \cite{steinbach2005cluster}, which optimizes cost functions in terms of data points and cluster centers.
Considering the CRP $ \bm{\zeta} $ used in (\ref{alpha}) and (\ref{gravity}), we name our model the Cluster Representative SBM (CRSBM).

\subsection{Model Parameters}
Let $ \bm{\beta} $ be the parameter of the node popularity function $ f $ and $ \vartheta = \{ \omega, \bm{\beta}, \bm{\zeta}\}  $  be the parameter set of CRSBM. Combining (\ref{gravity}) and $ p_{ij} = g_{ij}\omega_{z_i,z_j} $  with (\ref{Eq1}), we obtain the likelihood 
\begin{align}\label{likelihood}
P(G|z,\vartheta)  & = \prod_{i < j} (g_{ij}\omega_{z_i,z_j})^{a_{ij}}(1-g_{ij}\omega_{z_i,z_j})^{1-a_{ij}} \notag \\
& = \prod_{i < j}g_{ij}^{a_{ij}} \prod_{r \leq s} \omega_{rs}^{m_{rs}}e^{-\Xi_{rs}\omega_{rs}},
\end{align}
where the Poissonian approximation has been applied in the second equality. In \eqref{likelihood}, $ m_{rs} = \sum_{ij} \delta_{z_i,r}a_{ij}\delta_{z_j,s}/(1+\delta_{rs})  $ is  the number of edges in block $ (r,s) \in [q] \times [q]$, and $ \Xi_{rs} = \sum_{ij} \delta_{z_i,r}g_{ij}\delta_{z_j,s}/(1+\delta_{rs})  $, where $ \delta $ is the Kronecker delta.

It is common to assume that the  membership $ z $ of each node is independent due to the i.i.d. edges in SBM,  so the prior  of $ z $ can be chosen as a multinomial distribution ${\pi}(z) = \prod_i \nu_{z_i} $, where $ \nu_r $ is the possibility of any node $ i $ in community  $ r $,  satisfying the normalization  $ \sum_{r=1}^q  \nu_r =1$.  From the conditional probability formula $ P(G,z | \vartheta ) =   P(G|z,\vartheta) \pi(z) $, it follows that 
\begin{equation}\label{KEY}
P(G,z | \vartheta )= \prod_{i}\nu_{z_i} \prod_{i < j}g_{ij}^{a_{ij}} \prod_{r \leq s} \omega_{rs}^{m_{rs}}e^{-\Xi_{rs}\omega_{rs}}.
\end{equation}
Using the Lagrange multiplier method to maximize the logarithm $ \log P(G,z | \vartheta ) $ with respect to $ \nu_{r}  $ under the constraint $ \sum_{r=1}^q  \nu_r =1$,  we obtain that 
\begin{equation} \label{nu}
\nu_{r} = \frac{1}{n}\sum\nolimits_{i} \delta_{z_i,r}, ~ r \in [q].
\end{equation} 
Given the likelihood (\ref{KEY}),  for the parameter $ \omega $ that describes the block structure in $ G $,
the maximum likelihood estimation (MLE)  $ \partial \log  P(G,z|\vartheta) / \partial \omega_{rs} = 0 $ yields that
\begin{equation}\label{omega}
{\omega}_{rs} = \frac{ m_{rs} }{\Xi_{rs}}= \frac{m_{rs}(1+\delta_{rs})}{n_{r}^{s}n_{s}^{r}},
\end{equation}
where    $ n_r^s = \sum_i \delta_{z_i,r} f_{is} $  with    $ f_{is}$ being the abbreviation of $ f(\alpha_{is}) $  and $ n_s^r =  \sum_j \delta_{z_j,s} f_{jr}$. 
The estimation of  $ \bm{\zeta} $ and $ \bm{\beta} $ is relevant to the choice of the function $ f $, which will be discussed in Section~\ref{ms} in detail.

\begin{remark} \rm
	In the Bayesian view, one may choose a maximum entropy prior $ \pi(\omega) = \overline{\omega}^{-1}e^{\omega / \overline{\omega}} $  for $ \omega_{rs} $, where $ \overline{\omega} $  denotes the average of $ \omega $, and then the maximum a posteriori (MAP) estimation gives $ \omega_{rs} = m_{rs} / (\Xi_{rs}+ \overline{\omega}^{-1}) $ \cite{Peixotonp}. Note that  the average  linking possibility is $ \langle p \rangle = 2m/n^2 $, in DCSBM, $ \overline{\omega} = 2m/(c^2 n^2) =  O(n^{-1})$. Similarly, when the range of $ f(\alpha) $  is $ O(1) $, 
	$ \overline{\omega} $ is also  $ O(n^{-1}) $ and $ \Xi  $ is $ O(n^2/q^2) $ in CRSBM. Therefore, the MAP estimate of $ \omega $ is equivalent to the MLE in (\ref{omega})  when $ n \gg q^2 $.
\end{remark}

\section{BP Algorithm and Detectability}
In this section we first develop  an efficient algorithm to infer the community memberships based on Belief Propagation (BP), a classical framework for the estimation of marginals in probabilistic models \cite{ComputationPhysics2009}. And then we  investigate the detectability of communities  for  the proposed algorithm to clarify the contribution of  attributes {in the data fusion}, which is also  an  analysis  on  algorithmic effectiveness.

Before proceeding,  we   note that it is a common assumption in BP-based  methods that the  network $ G $ is sparse, that is, $ m = O(n) $ and $ p_{ij} = O(2m/n^2) = O(n^{-1})$. In words, it means that the number of edges $ m  $ is in the same order of the number of nodes  $ n  $.   In fact, it is also shown that BP algorithms also have  good performances on networks with   relatively large average degrees \cite{Zhang2014}. 

\subsection{BP Inference for CRSBM}

According to Bayes' rule, the posterior distribution of $ z $ follows $ P(z|G,\vartheta) = P(G,z|\vartheta) / \sum_z P(G,z|\vartheta) $, where $  P(G,z|\vartheta)  $ is shown in \eqref{likelihood}, and the possibility of each node $ i $ belonging to any community $ r $ is  $  P(z_i = r|G,\vartheta) = \sum_{z:z_i=r}P(z|G,\vartheta) $. To infer this marginal distribution, for each ordered pair $ (i,j)  \in V \!\times \! V, i\! \ne\! j $, BP defines \textit{messages} from $ i $ to $ j $, denoted by $ \psi_r^{i \to j}  $,  which means the   marginal of $ z_i = r $ conditioned on $ z_j $. Assuming that the distribution of the neighbors $ \partial i = \{j| a_{ij} =1 \} $  of  node $ i $ only correlates one another through $ i $, which implies that $ i $ and its neighbors approximately form a locally tree-like structure \cite{Decelle2011a,Moore2017},  the joint distribution of $ z_{\partial i} =  \{z_\ell | \ell \in \partial i\} $ conditioned on $ z_i $ is then the product of the marginals of $  z_{\partial i} $. In this case, $ \psi_r^{i \to j}  $ from $ i $ to $ j $  can be  recursively expressed by the messages from other nodes except $ j $ using the sum-product rule \cite{ComputationPhysics2009}. Based on the posterior distribution $ P(z|G,\vartheta) $, we derive the BP equation for the message $ \psi_r^{i \to j} $ as
\begin{align}\label{mess}
\psi_r^{i \to j} 
= \frac{\nu_{r}}{Z^{i \to j}} \prod_{l \notin \partial i} \left( 1\!-\! \sum\nolimits_s{ \psi_s^{l \to i}W_{sr}^{\ell i} }\right) \!\!\prod_{l \in \partial i \backslash j }\!\!\left( \sum\nolimits_s{ \psi_s^{l \to i}W_{sr}^{\ell i}}\right), 
\end{align}
where $ W_{sr}^{\ell i} = g_{li}\omega_{sr} $ is the coupling weight between groups $ s $ and $ r $, and $Z^{i \to j} $ is  the normalization factor with $ \sum_{r=1}^{q} \psi_r^{i \to j} =1 $. 
The marginal  of $ i $ can then be  estimated according to the messages that $ i $ receives, that is,  
\begin{align}\label{belief0}
\psi_r^i\! = \! \frac{\nu_{r}}{Z^i}\!\prod_{l \notin \partial i}\!\left( 1\!-\!  \sum\nolimits_s{ \psi_s^{l \to i} W_{sr}^{\ell i}  }\right) \! \prod_{l \in \partial i}\!\left( \sum\nolimits_s{ \psi_s^{l \to i} W_{sr}^{\ell i} }\right), 
\end{align} 
where $ \psi_r^i $ is the estimate of $   P(z_i = r|G,\vartheta) $, which is also referred to as    \textit{belief} in the BP algorithm. The main difference between  $ \psi_r^i $  and $ \psi_r^{i \to l}  $ is that whether the message from node $ l $ is included.  Note that in the case $ l \notin \partial i$, the  additional term in the product of $ \psi_r^l $ is $ 1-\sum\nolimits_s{ \psi_s^{l \to i}g_{li}\omega_{sr}} $, where  $ \sum\nolimits_s{ \psi_s^{l \to i}g_{li}\omega_{sr} }  = O(p_{li}) = O(n^{-1}) $ is sufficiently small with increasing $ n $. Then it follows that  $ \psi_r^{l \to i} = \psi_r^{l} + O(n^{-1}) $ and
$  1-\sum_s{ \psi_s^{l \to i}g_{li}\omega_{sr} }  \approx 1-\sum_s{\psi_s^{l}g_{li} \omega_{sr} }  \approx \exp(-\sum_s{ \psi_s^{l}g_{li}\omega_{sr}}) $. Therefore,  the message  $ \psi_r^{i \to j}  $  can be   written as
\begin{equation}\label{appr}
\psi_r^{i \to j}  = \frac{\nu_r}{Z^{i \to j}} e^{-h^i_r}\prod_{l \in \partial i \backslash j }\ \sum\nolimits_s{ \psi_s^{l \to i}(f_{lr}\omega_{sr}f_{is})}, 
\end{equation}
where
\begin{equation} \label{extfld}
h^i_r \triangleq \sum_l \sum_s{g_{li} \psi_s^{l}\omega_{sr}} = \sum_l \sum_s{ \psi_s^{l}f_{is}\omega_{sr}}f_{lr}, 
\end{equation} 
is the so-called auxiliary external field. Accordingly, the belief in (\ref{belief0}) can be  approximated as 
\begin{align}\label{belief}
\psi_r^i =  \frac{\nu_r}{Z^{i}} e^{-h^i_r} \! \prod_{l \in \partial i}\!\sum_s{ \psi_s^{l \to i}f_{is}\omega_{sr} f_{lr}}.
\end{align} 

As long as the function $ f $ and the parameter set $ \vartheta $ are given, the marginal $  P(z_i = r|G,\vartheta) $ can be inferred via iterating BP equations (\ref{appr}), (\ref{extfld}) and  (\ref{belief})  for each ordered node pair $ (i, j) \in \mathcal{E} \triangleq \{(i,j) \mid a_{ij}\!=\!1\}  $ until the convergence of $  \{\psi_r^i \}$. For clarity, we  present the detailed steps in advance in Algorithm 1 although the model learning procedure in Line~2 has not been discussed.

\begin{algorithm}[t]
	
	\SetKwInOut{Input}{\textbf{~~Input}}
	\SetKwInOut{Output}{\textbf{Return}} 

	
	\textbf{Input:} {$ G = (V,E,X) $,    number of communities $ q $}
	
	\textbf{Learning model:}  $ f $, $ \vartheta  = \{ \omega, \bm{\beta}, \bm{\zeta} \}$ 
	
	$  \psi_r^{i \to j} \!:=\! \text{rand}(0, 1)$, $ \psi_r^{i \to j}\!:=\!\psi_r^{i \to j}\!/\!Z^{i \to j}, \forall  (i,j) \in \mathcal{E}$\;
	
	get  $ f_{ir} $, $ \psi_r^i, h_r^i $ for  $ i \in V $, $ r\in [q]  $ by (\ref{gravity})(\ref{belief})(\ref{extfld})\;
	
	\While{beliefs $ \{\psi_r^i\} $ are not converged}{
		compute  $ \{h_r^i\} $ and store it into a $ n \times q $ matrix $ \mathcal{H} $\;
		
		set $ \Delta $ as a zero matrix of size $ q \times q $\;
		
		\ForEach{$ (i, j)  \in \mathcal{E}$ in random order}
		{
			$ h_r^\ell :=  \mathcal{H}_{\ell r} + \sum_{s=1}^q f_{\ell s} \Delta_{sr} $ for $ \ell\!\in\!\{i,j\} $\;
			
			update $ \psi_r^{i \to j} $, $ r\!\in\![q] $ by (\ref{appr})\;
			
			$  \bm{{\phi}}\! :=\! (\psi_1^{j},\ldots, \psi_q^{j}) $, update $ \psi_r^{j}  $ by (\ref{belief})\;
			
			$ \Delta_{rs} +\!\!=   (\psi_r^{j} -  \bm{{\phi}}_r)f_{js}\omega_{rs} $ for $ (r,s) \in [q] \times [q]$\;	 
			
		}  
	}
	
	\Output{ $ \{ \psi_r^i\} $, $ z_i := \arg \max _r\{\psi_r^i\}$, $ i \in V , r \in [q] $ }
	\caption{BP inference for CRSBM}
	\label{alg1}
\end{algorithm}

In Algorithm 1, to achieve the convergence of BP equations, an asynchronous update scheme is used, which means that the messages and beliefs are computed using the latest updated values available instead of the values at the last iteration,  as shown by the inner loop in  Lines~8--12. It is also notable that according to (\ref{extfld}), the update of   $ \psi_r^\ell $ of any node $ \ell $ will affect  the values of $ \{h_r^i\} $ of every node $ i $. To reduce the time complexity, instead of updating all the $ h_r^i $, $ i \in V $ after each computation of $ \psi_r^\ell $, we adopt a lazy update strategy \cite{madsen2008belief}, where $ h_r^i $ and $ h_r^j $ are only updated before the computation of message $ \psi_r^{i \to j} $. In detail, we first compute and store all the $ \{h^i_r\} $ before the inner loop (Line 6), and accumulate the changes caused by each update of $ \psi_r^\ell $ (Line 12) during the iteration. Therefore,  $ h_r^i $ and $ h_r^j $ can   be computed using the changes and the stored initial values (Line 9).

\begin{remark}\label{r2} \rm
	Setting  $ f $ as the constant function $1 $, we recover the BP equations for  the standard SBM, one of which about the message reads 
	\begin{equation}\label{recover}
	\psi_r^{i \to j}  = \frac{\nu_r}{Z^{i \to j}} e^{-h_r}\prod_{l \in \partial i \backslash j }\left(  \sum\nolimits_s{ \psi_s^{l \to i}\omega_{sr} }\right), 
	\end{equation}
	where $ h _r = \sum_l \sum_s{ \psi_s^{l}\omega_{sr}}$ is the external field. 
	Moreover, replacing $ f_{is} $ with $ k_i/c $ in (\ref{appr})--(\ref{belief}), where $ c $ is the average node degree, the BP equations for DCSBM are  recovered.
\end{remark}

\begin{remark}\label{r2.5} \rm
{Based on the BP framework, there are a number of variants to improve the efficiency of BP. However, most of them require that the coupling weight $ W_{sr}^{\ell i} $ can be reduced to a matrix irrelevant to node pairs, i.e.,  $ W_{sr}^{\ell i} = W_{sr} $ \cite{bp1,bp2}. This cannot be satisfied by our model that reads $ W_{sr}^{\ell i} = g_{\ell i}\omega_{sr} $. Therefore, we use the classical sum-product algorithm for model inference.}
\end{remark}
\subsection{Detectability of Community Structure}

Without loss of essence, community detection algorithms are usually theoretically analyzed based on a  \textit{symmetric} variant of SBM (SSBM) for simplicity \cite{Decelle2011a,Krzakala2013,Zhang2014}, in which all the planted communities have the same size $ n/q $, and $  m_{rs} $ only has two distinct values for all the $ (r,s) \in [q] \times [q] $, $ m_{rs} = m_{in} $ if $ r=s $  and $ m_{rs} = m_{out} $ otherwise. We further  
denote the  intra- and inter-community degrees by $ c_{in} = 2m_{in}/n $ and $ c_{out} =m_{out}/n$, respectively, and then the average degree  of the network is  $ c \!=\!  q^{-1} (c_{in}\!+\!(q\!-\!1)c_{out}) $.

For the SSBM, (\ref{recover}) has a  factorized fixed point (FFP) $\forall (i,j)\in \mathcal{E}$, $ \psi_r^{j \to i} = 1/q $, which is a trivial solution that implies the failure of community detection. 
The convergence at the FFP can be investigated via the   linear stability analysis, which is described by the first-order derivatives of messages in (\ref{recover}) and  the corresponding $ q \times q $ message transfer matrix $ T \equiv T^i, \forall i \in V $ with  the entry
\begin{equation}\label{transfer}
T^i_{rs} \triangleq \left. \frac{\partial \psi_r^{i \to j}}{\partial \psi_s^{l \to i}} \right|_{\text{\emph{FFP}}}.
\end{equation}
For a sparse graph $ G $, it was conjectured in \cite{Decelle2011a} and proved in \cite{Mossel2018} that, when the  parameters in  (\ref{transfer}) are in line with those of the SBM  generating $ G $,  the FFP is not stable with random perturbation $ \psi_r^{i \to j} = 1/q + \xi_r $ if 
\begin{equation}\label{key}
\tilde{c}\lambda^2_1(T) > 1, 
\end{equation}
and thus community memberships can be inferred efficiently via (\ref{recover}). In (\ref{key}), $\tilde{c}  = \langle k^2 \rangle / \langle k \rangle - 1  $ is  the average number of neighbors which each node passes messages to, i.e., the average excess degree   with $ \langle k  \rangle  $ being the mean degree and $ \langle k^2 \rangle  $   the mean-square degree. In particular, for ER networks, it follows that $  \tilde{c} = c $.   $ \lambda_1(T) $ is the largest eigenvalue of $ T $, which is often employed to describe the strength of community structure \cite{Banks2016}.
Both empirical experiments \cite{Decelle2011a} and theoretical studies \cite{Moore2017} have shown that a larger $ \lambda_1(T) $ leads to a better recovery of the planted communities under the  condition  \eqref{key}.

The  critical value at $ \tilde{c}\lambda^2_1(T) = 1 $ is referred to as the detectability limit  of community structure, or the Kesten-Stigum (KS) bound  \cite{Kesten1966}. Further researches show that the same bound is also shared by other methods including modularity optimization \cite{Zhang2014} and spectral clustering \cite{Krzakala2013}.

\subsection{Detectability Analysis for BP on CRSBM}
Besides the algorithmic effectiveness, it is notable that the  detectability condition \eqref{key} indeed quantitatively describes the contribution of node degrees and community strength on the detection task. Considering this, we preform the detectability analysis for our method to characterize the effect of node attributes on communities in CRSBM.

Based on the SSBM,  we start from the case that each node has a categorical attribute $ \mathbf{x}_i =  \varsigma _i  \in [ {q}] $ that indicates its community, which satisfies $ \| \mathbf{x}_i - \mathbf{x}_j \| \in \{0, 1\} $ and $ \alpha_{ir} \in \{0,1\} $. Setting $ f(1) > f(0) $, we find  that the  trivial solution $  \psi_r^{i \to j} = 1/q$, $ \forall (i,j) \in \mathcal{E}  $ is not the fixed point of (\ref{appr}) in this situation. Reducing  (\ref{appr}) according to the SSBM, we observe instead that 
\begin{equation}\label{fp}
\psi_r^{i\to j} = 
\left\{\begin{array}{ll}
{\gamma /(\gamma+q-1) } & {r = \varsigma _i}, \\
{1/(\gamma+q-1)}  & {r\neq \varsigma _i},
\end{array}\right.
\end{equation}
is a fixed point, where $ \gamma = f(1)/f(0) >1 $ describes the level of the dependence on node attributes. In contrast, without dependence on attributes, i.e., setting $ \gamma=1 $,  the trivial FFP $  \psi_r^{i \to j} = 1/q$ is then recovered.
Eq.~\eqref{fp} tells that given $ \gamma>1 $,  the detectability limit of communities vanishes so long as the attributes are indicative, that is, the memberships indicated by the attributes are better than random guess,  which is in line with the result in \cite{Zhang2014shui}.

However,  the available useful nodal information is rarely adequate  to identify  communities  in real-world networks. One collection of nodes with the same categorical attribute can contain multiple communities due to the inhomogeneous interactions  within the category  (e.g., the Amazon co-purchasing network) \cite{Fortunato2016}.  
A natural question  {that closely relates to data fusion} in this situation is:  

\textit{Are the  multiple communities within  the same category  detectable by the  BP algorithm, or merged into one community as indicated by the node attributes?}

With this problem in mind, we consider the following nested case: There are $ q^*  $ planted communities in the network generated by SSBM, each node of which is annotated by one attribute from $ \tilde{q} \ge 2 $ categories, and each category contains $ q_b = q^*/\tilde{q} \ge 2 $ modular groups, which are hereafter referred to as \textit{brother} communities  for brevity. The distance of each node to its own category is $ 0 $, and those to other categories are $ 1 $. We use $ z  \in z^\varsigma \triangleq $  $ \{ q_b\varsigma -q_b+1, q_b\varsigma -q_b+2,\ldots, q_b\varsigma \}, \varsigma  \in [\tilde{q}] $ to label the brother communities in category $ \varsigma $.  Without loss of generality, we set $ f(0) =1 $, and  denote the value of $ f(1) $ by $ \gamma $.  For this case, we  find a fixed point of (\ref{appr}) as
\begin{equation}\label{ffp}
\psi_r^{i\to j} = 
\left\{\begin{array}{ll}
{\gamma /(q_b \gamma +q^*\!-\! q_b) } & {r \in z^{\varsigma_i}}, \\
{1/(q_b \gamma +q^*\!-\! q_b) }  & {\text{otherwise}},
\end{array}\right.
\end{equation}
at  which $  \psi^i_r = \psi_r^{i\to j} $ according to \eqref{belief}. It is notable that  the modular structure within each category is unidentifiable at this fixed point. Thus, following the pioneering studies  \cite{Decelle2011a,Moore2017,Zhang2014} on detectability, we analyze the linear stability of (\ref{appr}) at the fixed point (\ref{ffp}) with the actual model parameters. Using (\ref{transfer}), we obtain the message transfer matrix $ T^i $ with 
\begin{equation}\label{Trs}
T_{rs}^i = \frac{\omega_{rs} f_{is} \psi^i_r}{\sum_u\omega_{ru} f_{iu} \psi_u^i } - \psi_r^i \sum_u\left( \frac{\omega_{us}f_{is}\psi^i_u}{\sum_v \omega_{uv}f_{iv} \psi_v^i }  \right),
\end{equation}
where  $  \psi^i_r = \psi_r^{i\to j} $ is applied. Writing (\ref{Trs}) into the matrix-vector form, we arrive that
\begin{equation}\label{T}
T^i = (I - \bm{\psi}^i\mathbf{1}^ \mathrm{T}) (\tilde{D}^{-1} \Psi^i ~\!\Omega~\!F^i),
\end{equation}
where $ I $ is a $ q^* \times q^* $ identity matrix, $ \mathbf{1} $  is an all $ 1' $s column vector, $ \bm{\psi}^i = (\psi_1^i, \psi_2^i, \ldots, \psi_{q^*}^i)^\mathrm{T} $, $ \Psi^i = \text{diag}(\bm{\psi}^i) $, $ \Omega = [\omega_{rs}]_{q^* \times q^*} $, $ F^{i} = \text{diag}(f_{i1},f_{i2},\ldots,f_{iq^*}) $  and $ \tilde{D} $ is a diagonal matrix with its $ r $th  diagonal entry being   the $ r $th row sum of  $  \Psi^i ~\!\Omega~\!F^i $.  To  solve the eigenvalues of $ T^i $, we next discuss the value of $ \omega_{rs} $ in (\ref{Trs}).  

With   $ f(0)=1 $, we obtain according to the MLE in (\ref{omega}) that $  \omega_{rr} =  c_{in}/n $. Note that in the message passing process,  for each community, its brothers are indistinguishable from other groups owing to the identical group sizes and random initial messages. Therefore, the values of $ \omega_{rs}, r \neq s $ in (\ref{Trs}) is equivalent to the average value of the MLE,  
\begin{equation}\label{wrs}
\omega_{rs} =  \langle  \omega \rangle_{r \neq s} = \frac{c_{out}\left[q_b-1 + \gamma^{-2}(q^*-q_b)\right]}{n(q^*-1)}, \forall r \neq s.
\end{equation}  
With  the matrix $ \Omega $ in (\ref{T}) obtained, for the leading eigenvalue $ \lambda_1(T^i) $ we have the following theorem: 

\begin{theorem}
	For each node $ i \in V $, the eigenvalues of $ T^i $ are all real values and the largest eigenvalue of each $ T^i $ shares the same value 
	\begin{equation}\label{lbd}
	\lambda_1(T^i) \!=\! \lambda _1(T) \!=\!  \frac{\omega_{in} - \omega_{out}}{\omega_{in}\! + \!(q^*\! - \! 1 \! - \! q_b)\omega_{out} \! + \! q_b \gamma^{-1} \omega_{out}}, 
	\end{equation}
	where $ \omega_{in} = c_{in}/n $ and $ \omega_{out} =  \langle  \omega \rangle_{r \neq s}  $ is shown  in (\ref{wrs}). 
\end{theorem}
\begin{IEEEproof}
	Please see the Appendix.
\end{IEEEproof}
Combining Theorem~1 and (\ref{key}), we   obtain the  condition under which the brother communities within the same category are detectable.  To show this result succinctly, let  $ \epsilon = c_{out}/c_{in} $ denote the ratio of inter- and intra-community degrees, and then the detectability condition is  
\begin{equation}\label{dc}
\epsilon < \epsilon^*_\gamma = \frac{\sqrt{\tilde{c}} -1}{ \eta ( q^* -q_b + q_b \gamma^{-1}  + \sqrt{\tilde{c}} -1 )  }, 
\end{equation}
with $ \eta = (q^*-1)^{-1} [q_b-1 + \gamma^{-2}(q^*-q_b)]  < 1$. Setting $ \gamma = 1 $ in (\ref{dc}), we obtain the detectability  of the BP equation (\ref{recover}) back for SSBM, i.e., $ \epsilon \! < \! \epsilon^*_1 = (q^*\!+\!\sqrt{\tilde{c}}\!-\!1)^{-1}(\sqrt{\tilde{c}}\!-\!1) $. Given  $ \gamma >1 $,  we have $ \epsilon^*_\gamma >  \epsilon^*_ 1 $, which shows that leveraging the node attributes, the condition in (\ref{dc}) is less strict than that for SSBM.  Moreover, it is notable that (\ref{dc}) in fact suggests that the proposed model and   algorithm can take  advantage of  both network topology, described by $ \epsilon $, and node attributes, described by $ \gamma $, to detect communities.


\section{Model Selection and Algorithm Details}\label{ms}

We have shown the major impact of the node popularity function $ f $ in (\ref{gravity}), highlighting the importance of the choice of $ f $ {in   the   model}.
In the existing community detection literature, multiple available models are often compared and selected according to some criteria including minimum description length (MDL) and Bayesian model selection \cite{Peixoto2014,Yan2014}. However, because of the diversity of node attributes, it is hard to determine their description length or specify a prior distribution without strong assumptions, especially for continuous attributes. 

To solve this problem, we present a novel  model selection scheme   for our CRSBM  based on the effect of  attributes on community detection,  which  can be quantitatively  described by the detectability. After determining the form of $ f $, we develop a  parameter estimation method that cooperates with  the BP inference, and then present the whole node attribute-aware community detection algorithm. 

\subsection{Bounds of the Node Popularity Function}

In the model \eqref{gravity}, the relative distance is $ \alpha_{ir} \in [0, 1] $.   Note that for either categorical or continuous attributes,  $  \alpha_{ir} = 1 $ means that $ \mathbf x _i $ is completely different from those in $ \mathcal{C}_r $.  Therefore,  a reasonable upper bound $ \gamma^*  = f(1) $ of the popularity function $ f $ can be studied based on  the analysis of    categorical attributed networks. To this end, we inspect the detectability condition (\ref{dc}) in terms of categorical attributes.

Note that the critical value $ \epsilon^* _\gamma $ in (\ref{dc}) in fact limits the ``strength'', or formally, the statistical significance \cite{Zhang2014}  of the detected communities, which is described by the ratio $ \epsilon = c_{out}/c_{in}  $. In this sense,  (\ref{dc}) shows that the  indicative attributes relax the condition and   make weaker communities with larger $ \epsilon $ detectable. On the other hand, it also means that the over-dependence  on attributes can cause the emergence of communities of no statistical significance and the over-split of modular networks. Therefore, the ratio $\gamma =  f(1)/f(0) $, which describes the level of dependence on attributes, should be limited.

In general, for assortative modular networks, it is required that $ \epsilon \!<\!1 $ in SBM to guarantee the significance of the planted communities. By contrast,  $ \epsilon^*_\gamma > 1$ in (\ref{dc}) may lead to the emergence of some disassortative   structure. To avoid this side effect, we have $ \forall q_b \ge 2, \epsilon^*_\gamma \le 1 $, which is reduced to $   {\epsilon}^*_\gamma  |_{q_b = 2} \le 1 $ since that $ \epsilon^*_\gamma $ decreases as $ q_b $ increases. Further, note that in the interval $ [1, +\infty), $ $ \epsilon^*_\gamma $ is a monotonically increasing function of $ \gamma $, and the critical value of $ \gamma $  is the maximum real-valued solution of
\begin{equation}\label{upbound}
  {\epsilon}^*_\gamma  |_{q_b = 2} = \frac{(q^*-1)(\sqrt{\tilde{c}} -1)}{   ( q^*-3  + 2 \gamma^{-1}  + \sqrt{\tilde{c}}   ) [1+\gamma^{-2}(q^*-2)] } = 1
\end{equation}
with $ q^* \ge 4 $, which can be simplified to a cubic equation. Analyzing the solution of \eqref{upbound}, we find that it is required that $ \tilde c>4 $ to ensure  $ \gamma^*>1 $. 

For the cases where \eqref{upbound} fails,  we here present an alternative method for the choice of $ \gamma $. {In community detection,  $ \lambda_1(T) $ is a central measure relevant to algorithmic performance \cite{Moore2017,Banks2016}.  It is clear  from the condition (\ref{key}) that a large $ \lambda_1(T) $ benefits the recovery of communities, and this is also verified by the empirical studies in \cite{Decelle2011a}.}  For simplicity, we investigate the contribution of  $ \gamma $ to $ \lambda_1(T) $ in  an extreme case based on SSBM, where the categorical attribute $ \varsigma_i $ of each node $ i $ indicates its community $ z_i $ correctly, i.e., $ \forall i,  \varsigma_i = z_i $.
In this situation, the transfer matrix $ T $ is in the same form of (\ref{Trs}) and has $ q $ real-valued eigenvalues with the largest one  
\begin{equation}\label{eq24}
\lambda_1(T) = \frac{\omega_{in} - \omega_{out}}{\omega_{in}\! + \!(q\! - \! 2 \! +\!\gamma)\omega_{out}  } = \frac{\gamma^2-\epsilon}{\gamma^2\!+\!(q\!-\!2\!+\!\gamma)\epsilon}, 
\end{equation}
which can be derived analogously by the method in Theorem~1. The derivative of $ \lambda_1(T) $ with respect to $ \gamma $ is  given by
\begin{equation}\label{derivative}
\frac{d\lambda_1(T)}{d\gamma} = 
\frac{ \epsilon [\epsilon + \gamma (2 q-2+\gamma)]} 
{\left[\epsilon (q-2+\gamma)+\gamma^2\right]^2} > 0,
\end{equation}
which approaches $ 0 $ with increasing $ \gamma $.  To ensure the contribution of attributes
to $ \lambda_1(T) $ and reduce the impact of noise on detected communities, we select $ \gamma^* $ at which point the growth rate of $ \lambda_1(T) $ is small enough, that is, 
\begin{equation}\label{mu}
\left. \frac{d\lambda_1(T)}{d\gamma} \right|  _{\gamma^*} = \mu \left. \frac{d\lambda_1(T)}{d\gamma} \right| _{\gamma=1}, 
\end{equation}
where $ \mu \in (0,1) $ is a hyper-parameter. Eq.~\eqref{mu} has an approximate solution $ \gamma^* \approx \mu^{-1/3} [1+(q-1)\epsilon ] ^{2/3} $.  
In practice, considering that in real-world networks,  intra-community  edges  are usually more than inter- ones \cite{radicchi2004defining}, we have $ c_{in} \ge (q-1)c_{out} $. Taking the corner case of  $ c_{in} = (q-1)c_{out} $, we obtain
\begin{equation}\label{upbound2}
\gamma^* \approx (4/\mu) ^{1/3}.
\end{equation}
Based on  the bounds above, we set $ \gamma^* $ as the minimum value of the solutions given by \eqref{upbound} and \eqref{upbound2}.


\subsection{Model Learning and Parameter Estimation}
The above analysis on the  two-sided effects of  node attributes has indeed suggested several rules for the  selection of   $ f $ in CRSBM.  (I). Without loss of generality, $ f(0) = 1 $. (II). $ f(1)>f(0) $ and $ f(1) $ should be a limited value that can be decided by (\ref{upbound}) and (\ref{upbound2}). Generalizing Rule~II to the distance $ x \in (0,1) $, we further have: (III).~For any two points $ x_1 $, $ x_2 $   satisfying $ x_1 > x_2 $, $ f(x_1) \ge f(x_2) $, and $ f(x_1) - f(x_2) $ should be small if $ x_2 $ is close to $ x_1 $, that is, formally, the derivative is limited, $ f'(x) \in [0, C] $. (IV). Under the condition of Rule~III,  $ f $ should be in a form that makes $ \omega_{in}/\omega_{out} $ as large as possible, which enlarges $ \lambda_1(T) $  according to (\ref{lbd}) and (\ref{eq24}) and thereby improve the algorithmic performance.

Taking these rules together, it is shown that an $ S $-shape curve is a good choice of $ f $,  e.g., a Sigmoid-like function 
\begin{equation}\label{sigmoid}
f(x) = (\gamma^*-1)/\left[1+\exp(-\beta_1 x + \beta_2)\right] + 1, \beta_1 > 0,
\end{equation}
with the range $ (1, \gamma^*) $, 
whose parameter set is denoted by $ \bm{\beta} = \{\beta_1, \beta_2 \} $. Note that the log-likelihood $ \log  P(G|z,\vartheta)  $ contains the summation of  $ O(n) $ terms in the form of $ - \log \sum_i f_{ir} $, and  maximizing such a non-convex objective with respect to $ \bm{\beta} $ is  expensive and sensitive to initialization.
We next propose a heuristic method for the estimation of $ \bm{\beta} $ and $ f $  to avoid the ill optimization issue. 

Before proceeding, we first give some preliminaries. For each node $ j $ and community $ r $,  $ f(\alpha_{jr}) $ is reasonable to be close to the lower bound $   1 $ if  $ z_j = r $, otherwise $ f(\alpha_{jr}) $ should be close to the upper bound $ \gamma^* $. Based on this intuition, for each point $ x $, we can update $ f(x) $ heuristically according to the marginals $ \mathcal{P}_x \triangleq \{ \psi_r^j  \mid (j,r) \textit{~s.t.~} \alpha_{jr} \in \mathcal{N}_x \} $ with the corresponding $ \alpha_{jr} $ falling into the neighborhood   $ \mathcal{N}_x =  (x\!-\!dx, x\!+\!dx) $ of $ x $. To this end, we define the measure 
\begin{equation}\label{df}
\varDelta_x \triangleq  \frac{2\langle \psi_r^j \rangle } {\langle \psi_r^j \rangle  + (q\!-\!1)^{-1}(1\!-\!\langle \psi_r^j \rangle ) } \!-\! 1,
\end{equation}
where $ \langle \psi_r^j \rangle $ is the average of the marginals in $  \mathcal{P}_x $. Noting that $ \varDelta_x $  satisfies that $ \varDelta_x > 0 $ iff $  \langle \psi_r^j \rangle > 1/q $ and  $ \varDelta_x < 0 $ iff $  \langle \psi_r^j \rangle < 1/q $, we update  $ f(x) $  by
\begin{equation}\label{update}
{f}(x) = f_{0}(x) + |\varDelta_x| \cdot (b - f_{0}(x)),
\end{equation}
where $ f_0(x) \triangleq (\alpha_{max}-\alpha_{min})^{-1} (\gamma^*\!-\!1)x\!+\!1 $ is the initial setting of $ f(x) $,   $ b = 1 $ if $ \varDelta_x > 0 $ and $ b = \gamma^* $ otherwise. In \eqref{update}, the  term $ b-f_0(x) $ guarantees that  $ {f}(x) $ is within the interval $ [1, \gamma^*] $ given that $ \varDelta_x \in [-1, 1] $.

In practice, we update $  f(x) $ on a finite set of samples $\mathcal S = \{(x,f_0(x))\} $ according to (\ref{update}), and  $ \bm{\beta} $ is then estimated by the Least Squares Method (LSM) to guarantee that Rule~III and Rule~IV are satisfied. In detail, for the function  $ f(\cdot) $ in (\ref{sigmoid}),  the estimation of $ \bm{\beta}$ given updated samples                                                                                                                                                                                                                                                                                                                                                                                                                                                                                                                                                         $ \{(x,y)\} $  with $ y = f(x) $ can be solved by the linear least squares estimation of $ \bm{\beta} $ on the transformed samples $\mathcal T = \{(\tilde{x}, \tilde{y})\} $, where $  \tilde{x} = -x $ and 
\begin{equation}\label{ls}
\tilde{y} = \log (\gamma^*-y) - \log (y-1) = \beta_1 \tilde{x} + \beta_2.
\end{equation}

Following \cite{Decelle2011a,Zhang2014}, we adopt an iterative learning scheme for the proposed model, that is, the parameters are updated based on the results of last iteration. The $ \delta_{z_i,r}\in \{0,1\} $  terms in   (\ref{nu}) and (\ref{omega}) are relaxed to the marginal $ \psi_r^i $, which improves the robustness of parameter estimation.  This relaxation gives
\begin{equation}\label{estimate}
\nu_r =\frac{1}{n} {\sum\nolimits}_{i} \psi_r^i \text{~~and~~} n_{r}^{s} = \sum\nolimits_{i}\psi_r^i f_{is}.
\end{equation}
Different from $ \nu_r $ and $ n_r^s $ that relate to one-node marginals only, $ m_{rs} $ in (\ref{omega}) involves two-nodes marginals $ P(z_i,z_j) $, that is, $ m_{rs} = \sum_{i<j} [P(a_{ij}\!=\!1,z_i\!=\!r,z_j\!=\!s)+P(a_{ij}\!=\!1,z_i\!=\!s,z_j\!=\!r)] $, where $ P(a_{ij}\!=\!1,z_i\!=\!r,z_j\!=\!s) = P(a_{ij}\!=\!1 | z_i\!=\!r,z_j\!=\!s) P(z_i\!=\!r,z_j\!=\!s)$.  In  BP, $ P(z_i\!=\!r,z_j\!=\!s)  $ is estimated as $ \psi_r^{i \to j} \psi_s^{j \to i} $ if  $ i $ and $ j $  are adjacent \cite{Zhang2014}. The estimate of $ m_{rs} $ can then be  written as 
\begin{equation} \label{mrs}
m_{rs} \! = \! \sum_{i<j}\frac{a_{ij}\omega_{rs}}{Z^{ij}} ( f_{is}f_{jr} \psi_r^{i \to j} \psi_s^{j \to i} \! + \! f_{ir}f_{js} \psi_s^{i \to j} \psi_r^{j \to i} ).
\end{equation} 
Denoting the numerator in (\ref{mrs}) by $ {\aleph}_{rs}^{ij} $, the normalization factor is $ Z^{ij} =  \frac{1}{2} \sum_{r}\sum_s \aleph_{rs}^{ij} $.

To estimate  $ \bm{\zeta} $   in \eqref{alpha}, we simplify the log-likelihood $ \mathcal{L} =   \log  P(G,z|\vartheta) $ to
\begin{equation}\label{loglikelihood}
\mathcal{L}  =  \sum _i \sum  _s \kappa_{is} \log f_{is}- \kappa_{is} \log  {n_{z_i}^{-1}}  \sum_{\ell: z_\ell=z_i} f_{\ell s} + C,
\end{equation}
where $ \kappa_{is} = \sum_j  a_{ij}\delta_{z_j,s} $ is the number of edges between the node  $ i $ and   group $ s $,   $ n_{z_i} = \sum_\ell \delta_{z_i, z_\ell} $  is the number of nodes in the group $ z_i $ and $ C $ is a constant irrelevant to $ f $ and $ \bm{\zeta} $. Applying the second-order Taylor's approximation to $ \mathcal{L} $ at the average value $ \bar{f}_{z_i,s} \triangleq {n_{z_i}^{-1}}  \sum_{\ell: z_\ell=z_i} f_{\ell s} $, we have
\begin{equation}\label{taylor}
\mathcal{L}  \approx L =  -\frac{1}{2} \sum\nolimits_{i}  \sum\nolimits_s \kappa_{is} \left(  {f}_{i s}/\bar{f}_{z_i,s}-1\right) ^2 + C.
\end{equation}
Solving  $ \partial  {L} / \partial \bm{\zeta}_s =0 $, we obtain
\[
\bm{\zeta}_s =   {  \sum\nolimits_i  \kappa_{is}  \rho_{is} w_{is} \mathbf{x}_i} {\big / } {  \sum\nolimits _i   \kappa_{is}   \rho_{is} w_{is} },
\]
where $ w_{is} =  \|\mathbf{x}_i-\bm{\zeta}_{r}\|^{-2} \alpha_{is}(1-\alpha_{is})   $ and  $ \rho_{is} = (f_{is}-\bar f_{z_i,s}) (f'_{is} \bar{f}_{z_i,s} -f_{is} \bar{f}'_{z_i,s})  $ with $ f' $ being the derivative of $ f $. Notice that $ \rho $ can be either positive or negative, which may result in an anomalous cluster center $ \bm{\zeta} $ that has large distances with all the $ \mathbf{x}_i $. Considering this, we further simplify $ \rho_{is} \propto  (f_{is}-\bar{f}_{z_i,s})^2 $ by approximating the derivative $ f'_{is} $ as a constant, which yields
\begin{equation}\label{zeta}
 \bm{\zeta}_s \!=\!   {\sum\nolimits_i  \kappa_{is}w_{is} (f_{is}\!-\!\bar{f}_{z_i,s})^2   \mathbf{x}_i} {\Big / } {  \sum\nolimits _i   \kappa_{is}w_{is}  (f_{is}\!-\!\bar{f}_{z_i,s})^2 },  
\end{equation}
where $ \kappa_{is} $ can be relaxed as $ \kappa_{is} = \sum_j a_{ij} \psi_s^j  $ and $ \bar f_{z_i,s} $ can be relaxed as  $ \bar f_{z_i,s} = (n\nu_s)^{-1} \sum_i \psi_s^i f_{is}$ based on the one-node marginals in BP.

\begin{remark}\label{r3} \rm
	In Remark 2, we have shown  that the derived BP equations can be transformed into  those for SBM and DCSBM by changing $ f_{is} $ into $ 1 $ and $ c^{-1}k_i $ respectively. These conversions are also applicable to \eqref{estimate} and \eqref{mrs} for parameter estimation. Furthermore, the node degrees can also be  integrated into our CRSBM together with attributes by replacing $ f_{is}  $ with $ c^{-1}k_i f_{is} $ in Eqs.~\eqref{appr}--\eqref{belief} for   inference,  and in Eqs.~\eqref{estimate}--\eqref{mrs} for parameter estimation.
\end{remark}

\subsection{Algorithm Details and Time Complexity}
Using the proposed model learning scheme, we present in Algorithm~2 the whole community detection procedure for attributed networks based on CRSBM. In  Algorithm~2, we initialize   $  \bm{\zeta}_r, r\in [q] $ using the famous initialization method for cluster centers in k-means++ \cite{david2007vassilvitskii}.  After the initialization, we conduct BP inference and parameter learning process iteratively using an Expectation Maximization (EM)-like framework (Lines~5--15), 
where the E-step for the latent group membership $ z $ is performed by the BP inference, and in M-step the parameters $ \vartheta $ are estimated by MLE.

\begin{algorithm}[t]
	
	\SetKwInOut{Input}{\textbf{Input}}
	\SetKwInOut{Output}{\textbf{Output}} 
	
	\Input{$ G = (V,E,X) $,    number of communities $ q $}
	
	initialize   $  \bm{\zeta} $  by center  initialization  in k-means++\;
	get $ \gamma^* $ by (\ref{upbound}) and (\ref{upbound2}) with $ \mu\! = \!0.05, \tilde{q} = \! q $\;
	$ f_0(x):= (\alpha_{max}-\alpha_{min})^{-1} (\gamma^*\!-\!1)x\!+\!1 $, $ \omega := qc/n $\; 
	$ \omega_{rr}\! :=\!\omega(1\!+\!\gamma^*)^{-1} \gamma^*$, $ \omega_{rs} \!:=\! \omega (1\!+\!\gamma^*)^{-1} $ by $ \gamma^* $ in \eqref{upbound2}\;

	\For{$ \tau := 0 $ \rm{\textbf{to}} $ \tau_{max}-1 $}{
		
		get  $ \{\psi_r^i\} $ and $ z_i $ by BP inference in Algorithm~1\;
		divide $ [\alpha_{min}, \alpha_{max} ] $ into $ N_s\!=\!10 $ grids uniformly, use the midpoints $ \{x_{k}\} $ of the grids to form  $\mathcal S $\;
		
		compute $ \{\varDelta_{x_k}\}_{k=1}^{N_s} $ by \eqref{df}, $ x_1  \! <\! x_2 \! <\! \cdots  \! <\! x_{N_s} $\;
		
		\If{$ \varDelta_{x_1} < 0 $  \rm{\textbf{and}} $ \varDelta_{x_2}<0 $}{
			update  $ \{\bm{\zeta}_r\} $ and $ \{\alpha_{ir}\} $ by  \eqref{zeta}, \eqref{alpha}\; 
			\textbf{goto} Line~15\;
		}
		
		update $ f(x_k) $ for $ \{(x_{k}, f(x_k))\}_{k=1}^{N_s} $ in $ S $ by (\ref{update})\;
		
		get $ T $ by (\ref{ls}) and conduct LSM on $ \mathcal T $ to get $ \bm{\beta} $\;
		
		update $ \bm{\zeta} $ by (\ref{zeta}), update $ \{f_{is}\} $ with new $ \bm{\beta}, \bm{\zeta} $\;
		
		update $\nu_r, n_{s}^r, n_r^s,  m_{rs},  \omega_{rs} $ by (\ref{estimate}), (\ref{mrs}) and (\ref{omega})\;	
	}
	compute the GN modularity $ Q $ for the resulting communities at each iteration\;
	\Output{$ \{z_i\} $ corresponding to the largest $ Q $ }

	\caption{\text{  Node Attributed Community {Detection}}}
	
	\label{alg2}
\end{algorithm}

It is difficult to specify a universal convergence threshold of EM for various network data due to the different correlation between graph structure and node attributes. As pointed by Newman et~al. in \cite{Newman2016a}, the EM algorithm with superfluous iterations may converge to poor solutions. Considering this, we run the iterations for $ \tau_{max} = 10 $ times, and use the GN modularity $ Q $   \cite{Girvan2002a} of the  partition at each iteration as a measure to select the results (Line~16), where 
\[
Q =\frac{1}{2 m} \sum_{i, j}\left(a_{i j}-\frac{k_{i} k_{j}}{2 m}\right) \delta_{z_{i}, z_{j}}.
\]
Despite that the ground truth community divisions of real-world   networks may not show the optimal modularity, it works well on selecting good results among the divisions produced at multiple iterations.

On  the choice of the sample set $ \mathcal S $ for LSM,  the interval $ [\alpha_{min}, \alpha_{max}] $ is divided into $ N_s \!=\! 10 $ grids of equal length $ 2dx $ and $\mathcal S $ is composed of  $ (x_k, f(x_k)) $ with $ x_k, k\in [N_s] $ being the midpoint of the grids.  To ensure that the popularity function $ f $ in the form \eqref{sigmoid} is non-decreasing, i.e., $ \beta_1 > 0 $, we skip the update of $ \bm{\beta} $ if   $ \varDelta_x < 0 $ for  the first two grids of  $ [\alpha_{min}, \alpha_{max}] $ (Lines 9--11), which  mostly occurs in the early iterations of Algorithm~\ref{alg2}. In the early stage, the update of $ f $ may cause a drastic change to the membership $ \mathbf{z} $, so stopping re-estimating $ \bm{\beta} $ and keeping updating $ \bm{\zeta} $ aim to obtain good CRPs of the inferred communities. In practice, we empirically find that $ \bm{\zeta} $ can reach good points quickly, and the update of $ \bm{\beta} $  seldom suspends  for three  successive iterations.

Finally, we discuss the time complexity of the proposed method. In Algorithm~\ref{alg2}, the initialization steps cost $ O(qnd) $ time. For the parameter learning procedure, updating $ \{m_{rs}\} $ takes $ O(q^2m) $ time operations,  updating $ \{\nu_r\}, \{n_{r}^{s}\}, \{\bm{\zeta}_s\} $ and $ f $   takes $ O(qnd) $ time, and conducting LSM to estimate $\bm{\beta} $ takes $ O(N_s^2) = O(1) $ time.  The BP inference  is conducted by Algorithm~1. In Algorithm~1, at each iteration, there are $ O(m) $ messages $ \{\psi^{i \to j}\} $  to update,   each of which is a $ q\times 1 $ vector (Line~10), and the update of $ \Delta_{rs} $ and $ h^{\ell}_r $, $ \ell \in \{i,j\} $ takes $ O(q^2) $  time operations for each $ \psi^{i \to j} $, and thus the time complexity of BP inference is $ O(q^2 m) $. Finally, calculating the modularity $ Q $ costs $ O(n) $ time.   {In conclusion, Algorithm~2  has a time complexity of $ O(q^2m + qnd) $ composed of two parts. The factor of $ O(q^2m) $ mainly resulting from the model inference procedure keeps in the same order of the computational complexity of BP  leveraging graph topology only \cite{Decelle2011a,Yan2014} and the other factor of $ O(qnd) $ scales linearly with the number and the dimension of attributes.}

\section{Experiments}
In this section, extensive experiments on both artificial and real-world networks are conducted to demonstrate the performance of our model and algorithm. Since   the community assignment is  still in serious dispute when the clusters of   attributes mismatch structural communities \cite{Peel2016}, there are currently no widely accepted artificial benchmarks for attributed networks.  Following \cite{Newman2016a,Hric2016b},   synthetic SBM graphs with categorical node attributes are only used to validate the detectability analysis for our algorithm, while   real-life networks with ground truth communities  are employed in the comparison between our method and   baselines.

\subsection{Verification on the Detectability Condition} 
To verify the detectability condition in \eqref{dc}, we generate a collection of SBM graphs with $ q^* = 4 $ communities of the same node size $ 5000 $ and set the number of categories $ \tilde{q} = 2 $. The synthetic graphs are all with the same average degree $ c = 4 $, while $ c_{in} $ and $ c_{out} $ vary in different networks. For convenience, we fix $ \gamma = f(1)/f(0) = 2 $. By \eqref{dc}, the critical value of  detectability  is  $ \epsilon ^*  =1/2 $. More intuitively, the corresponding  ratio of  internal degree is $ k_{in}/c = c_{in}/(cq^*)= 2/5 $. We show in Table~\ref{cm} the confusion matrices $\mathscr{M}\in \mathbb{R}^{q^*\times q^*}$  of BP inference on  three SBM graphs. The SBM-generated networks are  with $  k_{in}/c \in \{ 7/19, 8/20, 8/19\} $ respectively and $ \epsilon \in \{4/7, 1/2, 11/24\} $ accordingly, and we set $ C_1 $ ($ C_3 $) and $ C_2 $ ($ C_4 $) to be in the same category.  From the gray colored diagonal blocks in Table~\ref{cm} we can see that when $\epsilon \ge \epsilon^* $, the two brother communities with the same categorical attributes  are mixed into one in the detected community structure, which results in $ \mathscr{M}_{11} =  \mathscr{M}_{33} =0  $. In contrast, with $ \epsilon = 11/24<\epsilon^* $, BP inference finds two communities in each category, as shown by $ \mathscr M_{rr} >0, \forall r \in [q^*] $,  that is, the brother communities are detectable with $ \epsilon $ below the detectability limit.  
From the results on the above three SBM graphs,   the correctness of  the detectability condition   \eqref{dc}  for CRSBM is verified. 
{Additionally, we note that the detection accuracy is quite low because $ \epsilon $ is too close to $ \epsilon^* $ in the third setting. This phenomenon has also been observed in the experiments of artificial attributed graphs in \cite{Newman2016a}. In contrast, real-world networks usually have much lower $ \epsilon $ \cite{radicchi2004defining}, and our CRSBM is very effective in practice, as indicated by the experiments in Section~\ref{sec:comp}. }

\begin{table}[t]
	\centering
	\caption{Confusion matrices of BP on the   SBM graphs with $ \epsilon =4/7 > \epsilon^* $, $ \epsilon = 1/2 = \epsilon^* $, and $ \epsilon = 11/24 < \epsilon^* $.  $ C_1 $ ($ C_3 $) and $ C_2 $ ($ C_4 $),  are in the same category. Each element in the matrices are normalized  into $ [0, 1] $ by the division of $ n_0 $. DC: detected communities. GT: ground truth.}
		\vspace{-2mm}
	\begin{tabular}{|c|c|   c c | c c| } 
		\hline
		$ \epsilon  $	& \diagbox{\fontsize{7pt}{0}{GT}}{\fontsize{7pt}{0}{DC}}  & $ C_1 $ & $ C_2 $ & $ C_3 $ & $ C_4 $\\ 
		\hline
		\multirow{4}{*}{$\frac{4}{7}$}&	$ C_1 $ & \cellcolor{gray!25} 0 &  \cellcolor{gray!25}  0.6780 & 0.0196 & 0.3024 \\ 
		
		&	$ C_2 $ &\cellcolor{gray!25} 0	 & \cellcolor{gray!25}  0.6792 & 0.0182 & 0.3026\\
		\cline{2-2}	
		&	$ C_3 $ &  0.0028 & 0.2156  & \cellcolor{gray!25} 0 & \cellcolor{gray!25}   0.7816 \\
		
		&	$ C_4 $ &   0.0066  &   0.2144 &\cellcolor{gray!25}  0 &\cellcolor{gray!25}  0.7790 \\
		\hline  
		
	\end{tabular}
	\\
	\vspace{2mm}
	
	\begin{tabular}{|c|c|   c c | c c| } 
		\hline
		$ \epsilon  $&	\diagbox{\fontsize{7pt}{0}{GT}}{\fontsize{7pt}{0}{DC}}   & $ C_1 $ & $ C_2 $ & $ C_3 $ & $ C_4 $\\ 
		\hline
		\multirow{4}{*}{$\frac{1}{2}$}&	$ C_1 $ & \cellcolor{gray!25} 0 &  \cellcolor{gray!25}  0.7804  & 0.0356 & 0.1840 \\ 
		
		&	$ C_2 $ &\cellcolor{gray!25} 	0 & \cellcolor{gray!25} 0.7970  & 0.0306 & 0.1724\\
		
		\cline{2-2}
		
		&	$ C_3 $ & 0.2378 & 0.0646  & \cellcolor{gray!25} 0 & \cellcolor{gray!25}   0.6976 \\
		
		&	$ C_4 $ &  0.2318  &   0.0648 &\cellcolor{gray!25}  0 &\cellcolor{gray!25}  0.7034 \\
		\hline  
		
	\end{tabular}
	\\
	\vspace{2mm}
	
	\begin{tabular}{|c |c|   c c | c c| } 
		\hline
		$ \epsilon  $&	\diagbox{\fontsize{7pt}{0}{GT}}{\fontsize{7pt}{0}{DC}}   & $ C_1 $ & $ C_2 $ & $ C_3 $ & $ C_4 $\\ 
		\hline
		\multirow{4}{*}{$\frac{11}{24}$}&	$ C_1 $ & \cellcolor{gray!25} 0.0472 &  \cellcolor{gray!25}  0.7635  & 0.1107 & 0.0786 \\ 
		
		&	$ C_2 $ &\cellcolor{gray!25} 0.0416	 & \cellcolor{gray!25}  0.7557 & 0.1133 & 0.0893\\
		
		\cline{2-2}
		
		&	$ C_3 $ & 0.1168 & 0.1235  & \cellcolor{gray!25} 0.1000 & \cellcolor{gray!25}   0.6597 \\
		
		&	$ C_4 $ &  0.1067  &   0.1205 &\cellcolor{gray!25}  0.1016 &\cellcolor{gray!25}  0.6712 \\
		\hline  
		
	\end{tabular}
	
	\label{cm}
\end{table}

\subsection{A Real-world Case Study}

\begin{figure}[t]{
		\vspace{-4mm}
		\centering
		\subfloat[The Pubmed network]{
			\includegraphics[width=113.5pt]{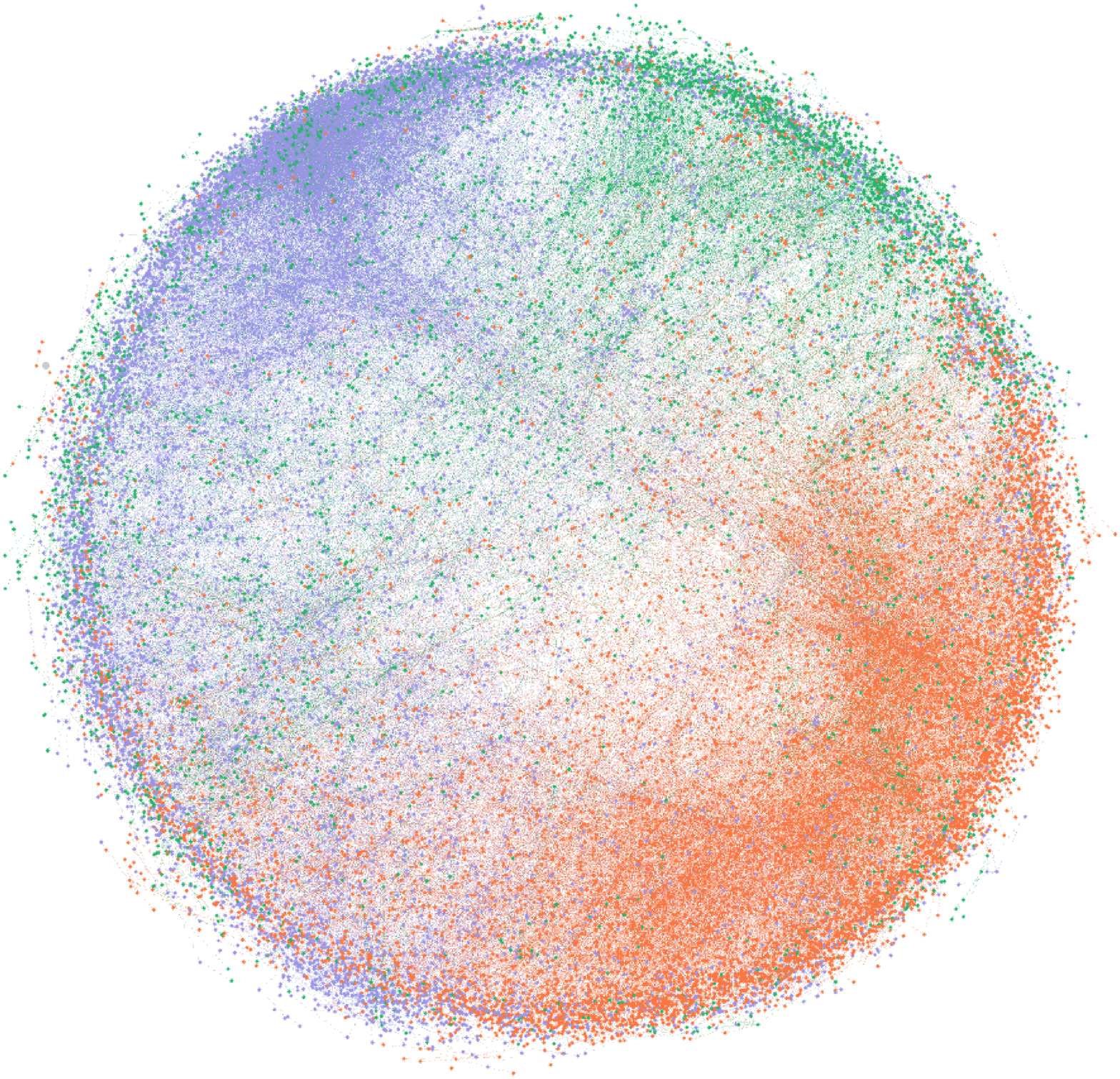}
		}
		\!\!\subfloat[Projected attributes and CRPs]{
			\includegraphics[width=126.5pt]{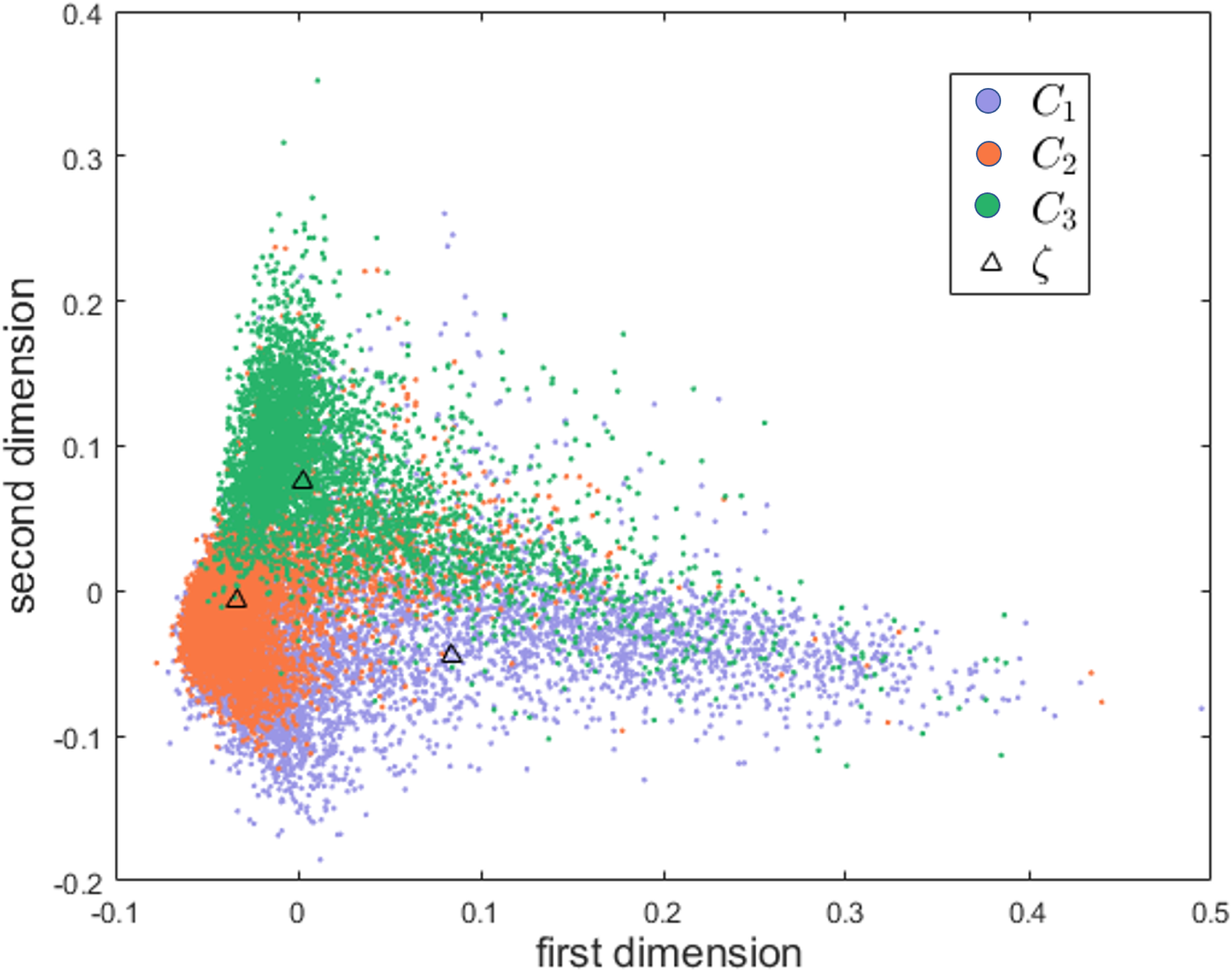}
		}
		\\
		\subfloat[The evolving $ f $ along with iterations]{
			\includegraphics[width=240pt]{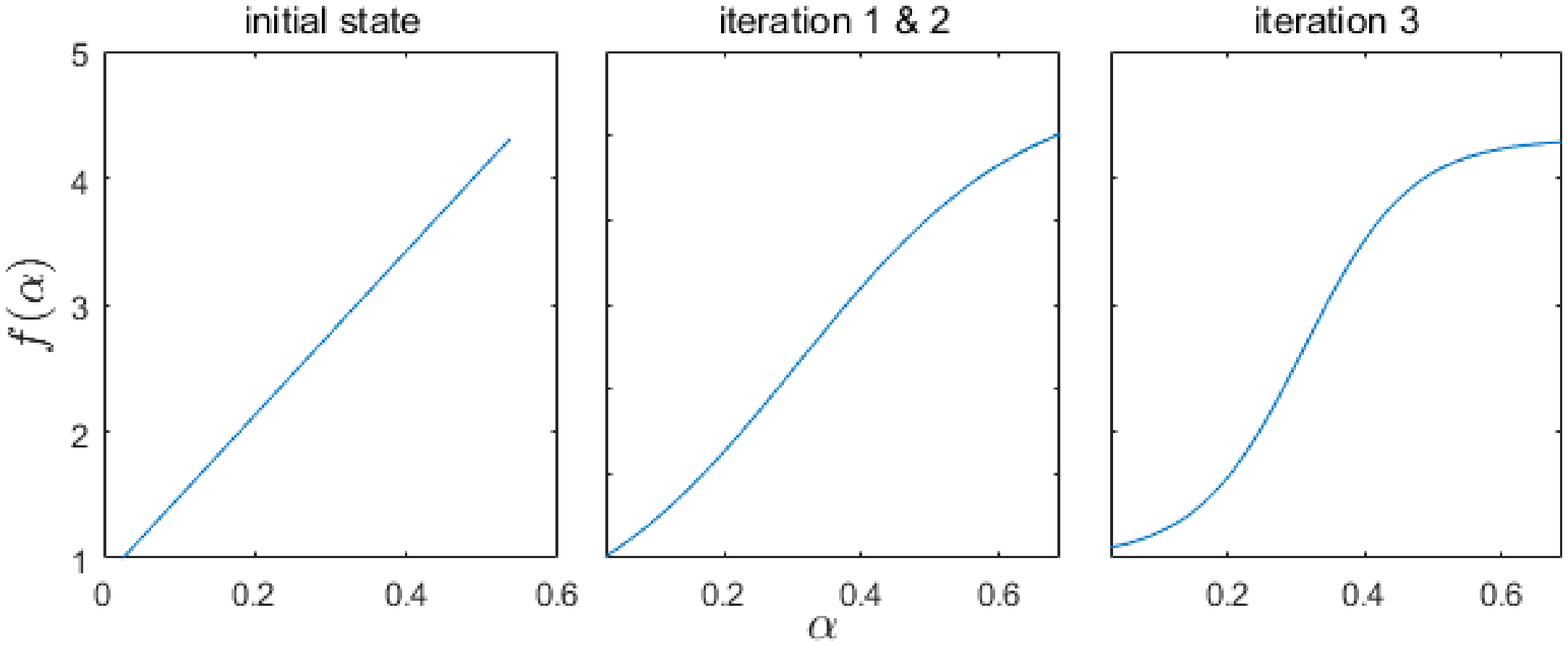}
		}
		\\
		\centering
		\subfloat[Detected communities in Pubmed]{
			\begin{minipage}[c]{1\linewidth}
				\centering
				\includegraphics[width=120pt]{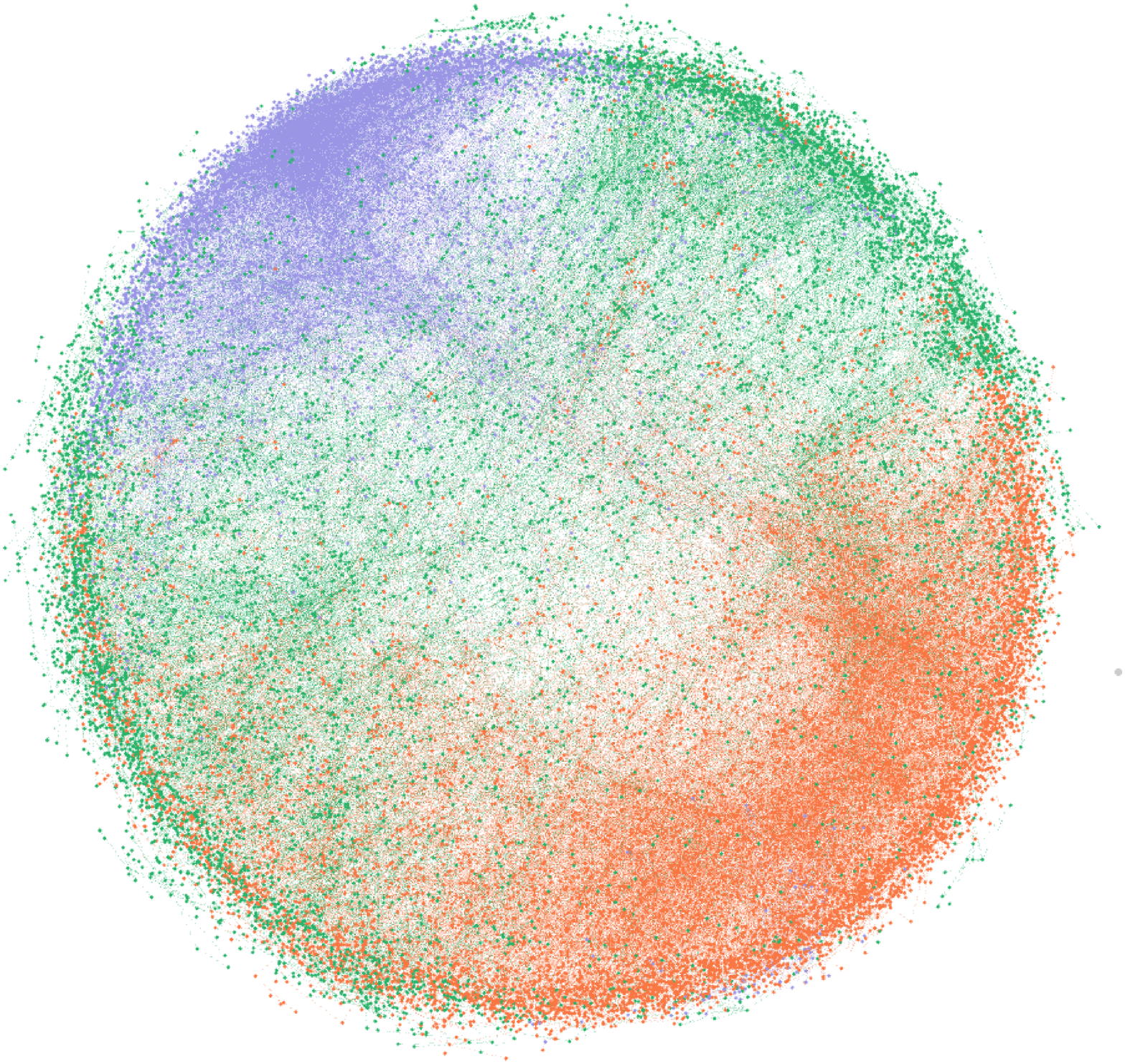}
			\end{minipage}
		} 
		\caption{Details of the detection process on Pubmed. (a). The ground truth communities in Pubmed are indicated by node colors. (b).  The projected data points of the estimated CRPs  and attributes in the ground truth communities $ C_1 $, $ C_2 $ and $ C_3 $. (c). At the second iteration, the conditions in Line~9 of Algorithm~\ref{alg2} are satisfied, and thus $ f $ is not updated. (d). The detected communities are shown with node position the same as (a).}
		\label{case}
		\vspace{-2mm}	
	}
\end{figure}

To illustrate our method in more detail, we here show the working process of Algorithm~\ref{alg2} via a case study on the citation network Pubmed, which contains 19729 nodes (papers), 44338 edges (citation relationships), 500 dimensional node attributes and 3 ground truth communities, as shown in Fig.~\ref{case}a.   The node attributes in Pubmed are sparse real vectors describing TF/IDF weights of words in the titles from a 500 word dictionary \cite{chunaev2020community}, whose first two principal components are visualized in Fig.~\ref{case}b via   principal component analysis (PCA)  \cite{abdi2010principal}. We can see from Fig.~\ref{case}b that a substantial portion of the attributes of each community mix with those belonging to other communities, which implies that mere attributes cannot indicate the communities well. 

Applying Algorithm~\ref{alg2} to Pubmed, the result at the third iteration shows the largest modularity  $ Q = 0.607 $ among $ \tau_{max} = 10 $ iterations, where the corresponding CRPs $ \{\bm{\zeta}_r| r =1,2, 3\} $ and the popularity function $ f $ are shown in Fig.~\ref{case}b and Fig.~\ref{case}c, respectively.  From the visualization, we observe that each   $ \bm{\zeta} $  locates at the position where the attributes in the same community are  densely distributed and the distances between different CRPs are relatively large. Therefore, the estimated $ \bm{\zeta} $'s are capable to be used as   cluster centers of attributes. Starting from the initial state of a linear function (Line 3, Algorithm~\ref{alg2}), the node popularity $ f $ changes into an $ S $-shape curve as the iterations proceed, which is in line with the model selection based on detectability analysis.

For the comparison with ground truth,  we present the detected communities in Fig.~\ref{case}d. It shows that our method estimates the group memberships of most nodes correctly, while the deviation is mainly caused by the nodes that have nearly the same amount of links to three communities, as shown by  the bottom-left of Fig.~\ref{case}a and Fig.~\ref{case}d. The quantitative evaluation  show that our method achieves the best performance   compared with the baselines on Pubmed, as will be  presented in Section~\ref{sec:comp}.

\subsection{Comparison with the Baselines}\label{sec:comp}
We further qualify the performance of the proposed method by comparing it with the baseline algorithms on   various real-life networks with ground truth available. The experimental settings are shown below.

\textit{Datasets}: Eight real-world network datasets are used in the experiments, including Citeseer, Cora, Pubmed\footnote{https://linqs-data.soe.ucsc.edu/public/}, Facebook, Twitter\footnote{http://snap.stanford.edu/},  Parliament\footnote{https://github.com/abojchevski/paican}, Arxiv and MAG \cite{mag}, whose profiles are summarized in Table~\ref{data}. For the datasets, two things need to be noted. First, Facebook and Twitter are two  collections of multiple social networks. Following \cite{joint,CHANG2019252,cde}, we use the one with largest node size in their collections respectively in the experiments.  Second, the node attributes in Pubmed, Arxiv and MAG are real-valued while others are binary-valued. The attributes of Pubmed are converted into binary ones due to the sparsity of the nonzero elements in the literature of SBM-based methods. In contrast, almost all of the node features of Arxiv and MAG are nonzero values.

\textit{Baseline algorithms}: Four classes of community detection methods are employed for  comparison. First, algorithms using graph structure only. To show the importance of fusing attributes in model-based approaches, we  adopt the extension of BP inference to DCSBM \cite{Yan2014} as a baseline, which can be derived from our algorithm according to Remarks~\ref{r2} and \ref{r3}.
Besides, the classical Louvain method  \cite{lou} and CommGAN \cite{gan}, a recently proposed approach based on deep learning, are compared in the experiments.  
Second, PGM-based algorithms incorporating both network topology and node attributes. Methods in this class include BAGC \cite{Xu2012},  CESNA \cite{Yang2013}, SI \cite{Newman2016a},  {BTLSC \cite{CHANG2019252},} and CohsMix3 \cite{Zanghi2010}.
{Third, network embedding approaches that describe   the relations among communities, graph structure and node attributes by linear and nonlinear mappings. In this line, NMF-based method ASCD \cite{ascd} and  NEC \cite{nec} based on graph neutral networks, are respectively included into comparison.}
{Additionally, in contrast to static optimization algorithms,  the methods based on the dynamic process of  networked systems are  also of  interest. To this end, we employ CAMAS \cite{BU201710} as a baseline, which is based on dynamics and the cluster properties in multi-agent systems.} Among all the baselines, only NEC can address arbitrary features, while CohsMix3 tackles Gaussian distributed attributes, and others  require categorical ones. 

The tuning parameters of all the baselines are set according to the authors' recommendations. For the statistical inference algorithms, we specify the ground-truth value $ K^* $ for the number of communities to be detected. Specially, there is a ground-truth cluster with only four disconnected nodes in Facebook. On this dataset, we set $ K^* \in \{8, 9\}$ respectively and report the best score.   
It is worth to note that SI \cite{Newman2016a} requires all the possible combinations of each dimension of node attributes, which is not scalable to  networks  in Table~\ref{data} that contain attributes of thousands of dimensions. To solve this problem, we first apply K-Means clustering \cite{david2007vassilvitskii} to the attributes, which converts the high-dimensional  feature  to univariate one,  and then use the clustering result as the input of SI. 
For CoshMix3 \cite{Zanghi2010} designed for continuous attributes, we conduct PCA on the binary feature vectors of and then take the real-valued  attributes in the projection space  as the input.

\begin{table}[t]
	\renewcommand\arraystretch{1.1}
	\centering
	\small
	\caption{Real-world Dataset Profiles}
	\vspace{-2mm}
	\begin{tabular}{l l l l l l l }
		\hline
		Class & Dataset & $ |V| $ & $ |E| $ & $ d  $ & $K^* $ & Attribute \\
		\hline
		Social & Twitter* & 171 &  796 &  578 &  6   & binary \\
		& Facebook* & 1045&  26749 & 576 & 9& binary\\
		Politics & Parliament & 451 &  5823 & 108 & 7 &binary\\	
		Citation &  Citeseer & 3312 &  4732 &  3703 & 6& binary \\
		& Cora & 2708 &  5429 & 1433 & 7 &binary \\         
		& Pubmed & 19729 &44338& 500 & 3 &real value \\
	 	&	Arxiv & 0.11M  & 1.3M  & 128 &  20 & real value \\
		 &  MAG   & 0.19M    &  3.4M & 128 & 9 &real value \\
			
		\hline
		\multicolumn{7}{l}{	$ K^*  $: \emph{Number of ground-truth communities} }{$ d $: \emph{Dimension of  attributes} }\\
		\multicolumn{7}{l}{\textit{$ M $: millions}}{\emph{Facebook*: network id: 107, Twitter*: network id: 629863}}\\
	\end{tabular}
	\label{data}
	\vspace{-3mm}
\end{table}

\textit{Evaluation metrics:} We adopt two widely used metrics in community detection to qualify the accordance between experimental results and ground truth and evaluate   the  competing methods, i.e., Average $ F_1 $ Score (AvgF1) and NMI \cite{Lancichinetti2009a}, whose definitions are as follows:
\[
\textit{AvgF}1\! =\! \frac{1}{2K^*} \!\!\sum_{C^{*}\! \in \mathscr{C}^{*}}\! \max_{C \in \mathscr{C}} \! F_1(C^*\!,C)
+
\frac{1}{2K} \!\!\sum_{C \in \mathscr{C}}\! \max_{C^{*}\! \in \mathscr{C}^{*}} \! F_1(C\!,C^*),
\]
\[
{\!N\!M\!I} = \frac{-2 \sum_{p=1}^{K} \sum_{q=1}^{K^{*}} n_{p q} \log \frac{n_{p q} n}{n_{p \cdot} n_{\cdot q} }}{\sum_{p=1}^{K} n_{p \cdot} \log \frac{n_{p\cdot}}{n}+\sum_{q=1}^{K^{*}} n_{ \cdot q} \log \frac{n_{ \cdot q}}{n}},
\]
where $ C\in \mathscr{C} $ is a  community detected by an algorithm, $ C^{*}\! \in \mathscr{C}^{*} $ is a ground-truth community, $ K $ is the number of detected communities, $ K^* $ is that of ground truth, and $ F_1(C_p, C_q) $ is the $ F_1 $ score between two sets $ C_p $ and $ C_q $. $ n_{pq} = |C_p\cap C_q |$, $ n_{p \cdot}  = \sum_q n_{pq}$ and $ n_{\cdot q} = \sum_p {n_{pq}} $.   By definition, higher NMI and AvgF1 scores indicate better community divisions. 

Note that CAMAS \cite{BU201710} and CESNA \cite{Yang2013}   may discard anomalous nodes in the detection procedure. Consequently, the NMI index that requires the compared partitions to cover the same node set is unable to evaluate the performances of these two baselines. Instead, we use the  extension of NMI named ONMI  in \cite{Lancichinetti2009a} for overlapping community detection as the evaluation metric. 

\begin{table*}[htb]
	\renewcommand\arraystretch{1.2}
	\centering
	\caption{{Comparison of the AvgF1 and NMI/ONMI Scores between Our CRSBM and Baselines on Binary Attributed Networks}}	
	\vspace{-3ex}
	\begin{tabular}{l|| c| c || c |c ||  c |c || c |c || c| c||  c| c}
		\hline
		Network &  \multicolumn{2}{c||}{Twitter*} &   \multicolumn{2}{c||}{Facebook*} &  \multicolumn{2}{c||}{Cora} &  \multicolumn{2}{c||}{ Citeseer}  &  \multicolumn{2}{c||}{Pubmed} &  \multicolumn{2}{c}{Parliament}   \\
		\hline
		Metric \% & AvgF1 & ~~NMI~ & AvgF1 & ~~NMI~ & AvgF1 & ~~NMI~& AvgF1 & ~~NMI~& AvgF1 & ~~NMI~& AvgF1 & ~~NMI~ \\
		
		\hline
		DCSBM           &   49.33   &   55.47  &     38.73   &   43.23    &  53.50   &  36.96   &    39.17    &            16.34    &    55.33     &  18.14    & 51.23    & 41.96          \\
		commGAN\!& 47.63 & 33.99 & 32.06 &  26.79   & 31.78&6.72  &  25.03& 5.90 &  41.47 & 0.11 & 51.00 &20.91 \\
		Louvain  &37.04    &  54.64     &  35.30  &    55.82     &56.42    &  43.31       &    41.53  & 27.74          &  35.11  &  17.66     &   55.78 &  70.53  \\

	\hline
	\hline
	
		BAGC        &     N/A     & N/A     &    27.67    &  9.14   &    36.46 & 16.97    &   N/A     & N/A     &   36.33     & 8.31     &   29.76  &    5.27        \\
		
		SI      &  50.89     & 54.52    & {51.61}& {57.80}    &  49.50   &  36.08    &  42.33  & 28.13    &  43.17   & 9.67     & 43.90    & 63.53      \\
		
		{BTLSC} & 56.91 & {66.52} & 43.54 & 56.42 & 46.61 & 32.04 & 34.12 & 15.70 & 56.91 & 17.69 &  62.38 & 69.74 \\

		CohsMix3 &   27.07     &  5.56         &  14.85    & 10.52     & 17.74    & 4.92     &       19.83 &      3.38      &  33.63        & 0.01     & 32.49    & 3.12       \\
		NEC & 48.80    & 42.31  & 44.11   & 40.21&  36.50   & \textbf{55.84*}      &   30.67   &27.83*
		& 46.17   & 4.96  &58.14 & 58.05\\
		ASCD & 57.75&\textbf{66.89} & 45.13 & \textbf{58.41} & 51.35 & 35.56 & 40.42 & 24.96 & 50.83 & 14.85&66.53&74.77 \\
		
		\textbf{CRSBM}       &   \textbf{58.96}    & {59.31}  &\textbf{56.77}  &53.96  &  \textbf{57.93}   &   {44.42}     &  \textbf{48.03} & \textbf{29.12}            &   \textbf{62.98}       &   \textbf{25.73}   & \textbf{72.21}    &  \textbf{78.65}         \\

		\hline
		\multicolumn{13}{c}{~}\\  
		\hline
		Metric \% & AvgF1 & ONMI& AvgF1 & ONMI& AvgF1 & ONMI& AvgF1 & ONMI& AvgF1 & ONMI& AvgF1 & ONMI \\
		
		\hline
		CAMAS             &   34.02       & 17.93   &  31.94      &  \textbf{38.42}	  &  8.94   & 0.01    &   5.80     & 0.01           &     8.48     & 0.01     & 40.94    &   34.46      \\
		CESNA     &    43.72     & 15.53  &  {49.05}  &     {27.02}  &  46.14    &    19.80  &  3.38    & 2.26          &       22.08   &   1.01 & 65.64    & 49.58         \\
		\textbf{CRSBM}      &   \textbf{58.96}    & \textbf{31.15} &\textbf{56.77}  &32.90  &  \textbf{57.93 }  & \textbf{27.61}   &  \textbf{48.03} & \textbf{12.25}           &   \textbf{62.98}       &   \textbf{19.72}   & \textbf{72.21}    &   \textbf{57.02}         \\
		\hline
		\multicolumn{13}{l} {\makecell[l]{
	* These two results are directly drawn from original paper of NEC \cite{nec}, while other scores are reported according to our reproduction. \\ 
    Based on our implementation, the NMI on Cora is 20.23\%, and that on Citeseer is 13.57\%.}
	}\\

	\end{tabular}
	\vspace*{-3ex}
	\label{result1}
\end{table*}

The experiments were conducted on the datasets in Table~\ref{data}. We show the results on binary attributed networks (Pubmed included) in Table~\ref{result1}, and those on networks with real-valued attributes in Table~\ref{result2}, respectively, where the best scores for each network are highlighted in bold and N/A means that the algorithm only detected one trivial community on the network. { 
The experiments of NEC and commGAN were conducted on a NVIDIA RTX3090 GPU with 24GB GPU memory and  others  on a PC with Intel i9-10900X@3.7 GHz CPU and 128GB memory.}
\begin{table}
	
	\centering
	\renewcommand\arraystretch{1.2}
	
	\caption{{Comparison of Clustering Accuracy on Twitter* and Facebook*}}	
	\vspace{-3ex}
	\begin{tabular}{l|  c| c |  c |c }
		\hline
		Method & SI & BTLSC & ASCD &CRSBM \\
		
		\hline
		Twitter* & 0.4528  &  \textbf{0.6288}& 0.5682 & 0.5814\\
		Facebook*   & 0.3292  &  0.6548  &  0.4745    & \textbf{0.7042}  \\
		\hline
	\end{tabular}
	\vspace{-2mm}
	\label{result1-1}
\end{table}

From Table~\ref{result1}, we observe that:
First, our CRSBM is the only attributed community detection method that is superior to DCSBM on all the eight datasets, which shows that CRSBM can effectively {fuse}  attributes to improve the performance of detection. Second, CRSBM, BTLSC, SI, ASCD and NEC are effective on both dense and sparse networks, while CohsMix3, CAMAS and CESNA show  inferior performances on the networks that have a small average node degree  around $ 4 $. Third, our method significantly  outperforms the baselines on   Citeseer,  Pubmed, and Parliament. And our results on  Cora show the best score in terms of AvgF1 and the second best in terms of NMI. 

It is noticed that the ranks of competitors given by NMI and AvgF1 have remarkable difference  on Twitter* and Facebook*. In order to make the comparison more convincing, we additionally compare the clustering accuracy ($ AC $) scores of SI, BTLSC, ASCD and our CRSBM on these two datasets, which are displayed in Table~\ref{result1-1}. The clustering accuracy is defined as
\[
AC(C, C^*) = \frac{1}{n} \sum\nolimits_{i=1}^n \delta(C^*_i, map(C_i)),
\]
where $ map(\cdot) $ is the permutation that maps each label $ C_i $ to the equivalent label from the dataset.
It is shown in Table~\ref{result1-1} that CRSBM gives the highest accuracy on Facebook* and the second highest score on Twitter* only after BTLSC, while the accuracies of SI are the lowest on both datasets. Taking the three  metrics together, CRSBM achieves the best scores in terms of two on Facebook* and shows a competitive performance on Twitter*.

\begin{table}
	
	\centering
		\renewcommand\arraystretch{1.2}

	\caption{{Comparison between Baselines and CRSBM on Networks with Real-valued Node Attributes}}	
\vspace{-3ex}
	\begin{tabular}{l|| c| c || c |c }
		\hline
		Network &  \multicolumn{2}{c||}{Arxiv} &   \multicolumn{2}{c}{MAG} \\
		\hline
		Metric \% & AvgF1 & ~~NMI~ & AvgF1 & ~~NMI~ \\
		
		\hline
		DCSBM & 17.85 & 19.16 & 25.07 & 24.62\\
		Louvain   & 22.24    &  24.90       &       27.38      & 28.01   \\
		commGAN  & 21.73  & 11.38   & 30.33   & 20.93\\
		CohsMix3  & \multicolumn{2}{c||}{Out of Memory} &  \multicolumn{2}{c}{Out of Memory}   \\
		NEC  & \multicolumn{2}{c||}{Out of Memory} &  \multicolumn{2}{c}{Out of Memory}   \\
		\textbf{CRSBM}      	& \textbf{ 26.08}   & \textbf{ 24.95 }      & \textbf{ 38.83  }  & \textbf{ 32.13 } \\
		\hline
		
\end{tabular}
\vspace{-2mm}
\label{result2}
\end{table}

Finally, we report the experimental results on large networks. As indicated in Table~\ref{result2}, CRSBM beats the baselines on two large networks. For these datasets, CoshMix3 and NEC ran out of the memory of our device because of their $ O(n^2) $ space complexity. In contrast, our method only consumes $ O(qm) $ space and thus can be applied to large-scale sparse networks.
Overall, our method achieves the best performance among the competing approaches. Moreover, compared to other algorithms,   CRSBM also shows  better applicability to various node attributed networks, whose   edges may be sparse or dense, and node attributes may be categorical or real-valued.

\subsection{Comparison of Time Efficiency}
\begin{figure}[h]{
 		
 		\centering
 		\includegraphics[width=160pt]{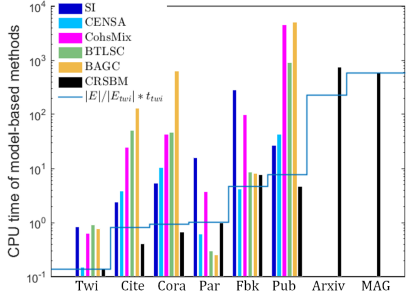}
 		\vspace{-2mm} \small
 		\caption{CPU times  in seconds  of the model-based algorithms. The CPU times are shown by   histograms, and the comparison between the increase of edge sizes and that of CPU time is shown by  blue stairs, where $|E| $ is the edge size of the dataset,  $ |E_{twi}| $ is that of Twitter,  and $ t_{twi} $ is the CPU time of our CRSBM on Twitter*.  Twi: Twitter, Par: Parliament, Cite: Citeseer, Fbk: Facebook*, Pub: Pubmed. Implementation: CRSBM, SI, CESNA in C/C++; BAGC, BTLSC in Matlab; CohsMix3 in R.}
 		\label{time}
 		
 }\end{figure}

For a  clear comparison on   time efficiency, we first present the  time complexity of the employed competing methods in Table~\ref{table4}. It shows that our algorithm has a competitive theoretical time efficiency compared to other baselines. {To validate this, we report the CPU times  of the  attributed community detection algorithms in Fig.~\ref{time}. It is noted that the GPU times of NEC are not included, because the main limit of the scalability of NEC is the memory usage.   We also compare the increase of CPU times (histograms) with that of the number of edges (blue stairs) on different datasets to show the time scalability. 
As displayed in Fig.~\ref{time}, CRSBM not only demonstrates the best time efficiency compared to the baselines, but also has a good time scalability on large networks with millions of edges.}

\begin{table*}[t]
	\renewcommand\arraystretch{1.1}
	\vspace{-2mm}
	\caption{  {Comparison of the Time Complexity between Our CRSBM and Attributed Community Detection   Baselines
	}}
	\vspace*{-2mm}
	\centering
	\begin{tabular}{c |c c c c c c c c}
		\hline
		Methods & CRSBM &  BAGC   & SI   & BTLSC  & CohsMix3   & CESNA  & NEC    & CAMAS     \\
		\hline
		Time complexity & 	$ O(q^2m+qnd) $   & $ O(q^2n^2)  $ &$ O(q^2m+2^d) $  &  $ O(q^2 n^2 + nd) $ & $ O(qn^2d) $  & $ O( m + qn ) $ &  $ O(n^2+qn) $ &  $ O(n^2) $\\
		\hline
		
	\end{tabular}
	\label{table4}
\end{table*}

\section{Conclusion}
{In this paper, we have proposed a novel  PGM named CRSBM  for attributed community detection  without any requirements on the distribution of attributes, which can be either categorical or real-valued. This mainly contributes to the incorporation of the distances between  attributes in the model}.  In detail,  we have first described the impact of attributes on node popularity by attaching a  function of the distances  to the classical SBM. Then to choose an appropriate node popularity function, which inherently relates to the model selection problem, we  analyzed the detectability of communities for CRSBM.  And it came out that a function showing an $ S $-shape curve is a good choice to describe the  popularity, {as well as the weight of different attributes in data fusion}. After that, an efficient algorithm was developed to estimate the parameters and detect the communities. Extensive   experiments on real-world networks  has shown that our method is superior to the  competing approaches.

For quantitative analysis, we have derived the  detectability condition for CRSBM, which has been verified by numerical experiments on artificial networks. As a quantification of the effect of node attributes on community detection,  the detectability shows that if there are multiple (but not all) communities with all their nodes  containing the same categorical attribute, the detectability can still be improved compared to that with attributes being ignored, where the improvement is mainly determined by the average node degree   as well as the level of the dependence on attributes.  

{In the future, we plan to apply our detectability-based model selection scheme to other methods for the comparison and choice of various \textit{a priori}  models and parameters.}

\appendix


\textit{Proof of Theorem~1:} For any two matrices $ T^{i} $ and $ T^{j} $ defined in \eqref{T}, it follows that $ T^i=T^j $ if $ \varsigma_i=\varsigma_j $, that is,  $ i $ and  $ j $ have the same categorical attribute. Otherwise, let $ z_i = r $ and $ z_j =s $, $ T^i $ can be  transformed into $ T^j $ by first swapping its $ r $th and $ s $th rows and then swapping the $ r $th and $ s $th columns, which are elementary transformations. Therefore, the matrices $ \{T^i|i \in V \} $ are similar to each other, and share the same eigenvalues. 

Note that $ \sum_{r=1}^{q^*}  \psi^i_r = 1 $, which yields $ \mathbf{1}^\mathrm{T}( I - \bm{\psi}^i\mathbf{1}^ \mathrm{T} ) = \mathbf{0}^\mathrm{T}$. Then it follows that $ \mathbf{1}^\mathrm{T} T^i = \mathbf{0}^\mathrm{T} = 0 \mathbf{1}^\mathrm{T} $. Thus, $ 0 $ is an eigenvalue of $ T^i $. 
Before solving for other eigenvalues of $ T^i $, we first present some notations. 
Let $ \mathbf v_{rs} \triangleq (0, \ldots, 1,0 \ldots,-1, \ldots,0)^\mathrm T$, where $ 1 $ is the $ r $th and $ -1 $ is the $ s $th entry, $ r \ne s $, while other entries are all $ 0 $. We also define an auxiliary matrix $ \tilde{T}^i \triangleq \tilde{D}^{-1} \Psi^i ~\!\Omega~\!F^i $, which satisfies that $ T^i \mathbf v _{rs} =    \tilde{T}^i \mathbf v _{rs} $.

Without loss of generality, let $ z_i = r = 1$, then $ F^i = \text{diag}(1, \ldots, 1, \gamma, \ldots, \gamma) $ with $ 1 $'s being the first $q_b $ entries,  and  $ \bm{\psi}^i \propto (\gamma, 1, \ldots, 1) $ with  $ \gamma $  the first entry. After some lines of linear algebra, we obtain that
$ \mathbf v_{1s}, s=2,\ldots,  {q}_b $ are $  q_b -1 $ eigenvectors of $ \tilde{T}^i $ with the corresponding eigenvalues sharing the same value 
\begin{equation}\label{a1}
\lambda _{1s}(\tilde{T}^i) = \frac{\omega_{in} - \omega_{out}}{\omega_{in}\! + \!(q^*\!+\! 1 \! - \! q_b)\gamma \omega_{out} \! + \!( q_b-1) \omega_{out}}.
\end{equation}
Similarly, setting $ r = q_b+1 $, we obtain that $ \mathbf v_{rs}, s \!=\! r\!+\!1, \ldots, q^* $ are $ q^*-q_b +1 $  eigenvectors of  with the corresponding eigenvalues sharing the same value 
\begin{equation} \label{a2}
\lambda _{q_b+1, s}(\tilde T ^i) \!=\!  \frac{\omega_{in} - \omega_{out}}{\omega_{in}\! + \!(q^*\! - \! 1 \! - \! q_b)\omega_{out} \! + \! q_b \gamma^{-1} \omega_{out}}.
\end{equation}
Given that  $ T^i \mathbf v _{rs} =    \tilde{T}^i \mathbf v _{rs} $, the values in \eqref{a1} and \eqref{a2} are  also  eigenvalues of $ T^i $. Now we have   found $ q^*-1 $ real eigenvalues of  $ T^i $. All the $ q^* $ eigenvalues of $ T^i $ are real since the complex eigenvalues must be conjugate.  The remaining one, denoted by $ \lambda_{last}(T^i) $, can be computed according to the fact that $  \sum_k \lambda_k(T^i) = \text{trace}(T^i)$, where $ \text{trace}(T^i) = \sum_{r}T^i_{rr} $ is the trace of  $ T^i $. 
Given that $ \gamma > 1 $ and $ \omega_{in} > \omega_{out} $, we have  $\lambda _{q_b+1, s}(T^i) > \lambda _{1s}(T^i) > 0  $, and by direct computation we also find that  $ \lambda_{last}(T^i) <  \lambda _{q_b+1, s}(T^i) $. Therefore, $ \lambda _{q_b+1, s}(T ^i) $ in \eqref{a2} is the largest eigenvalue among all the $ q^* $ real eigenvalues of $ T^i $, $ \forall i \in V $. This completes the proof.

\section*{Acknowledgments}
The first author would like to thank Xiaowei Zhang in UESTC for helpful discussions.


\vspace*{-5ex}
\begin{IEEEbiography}[{\includegraphics[width=1in,height=1.25in,clip,keepaspectratio]{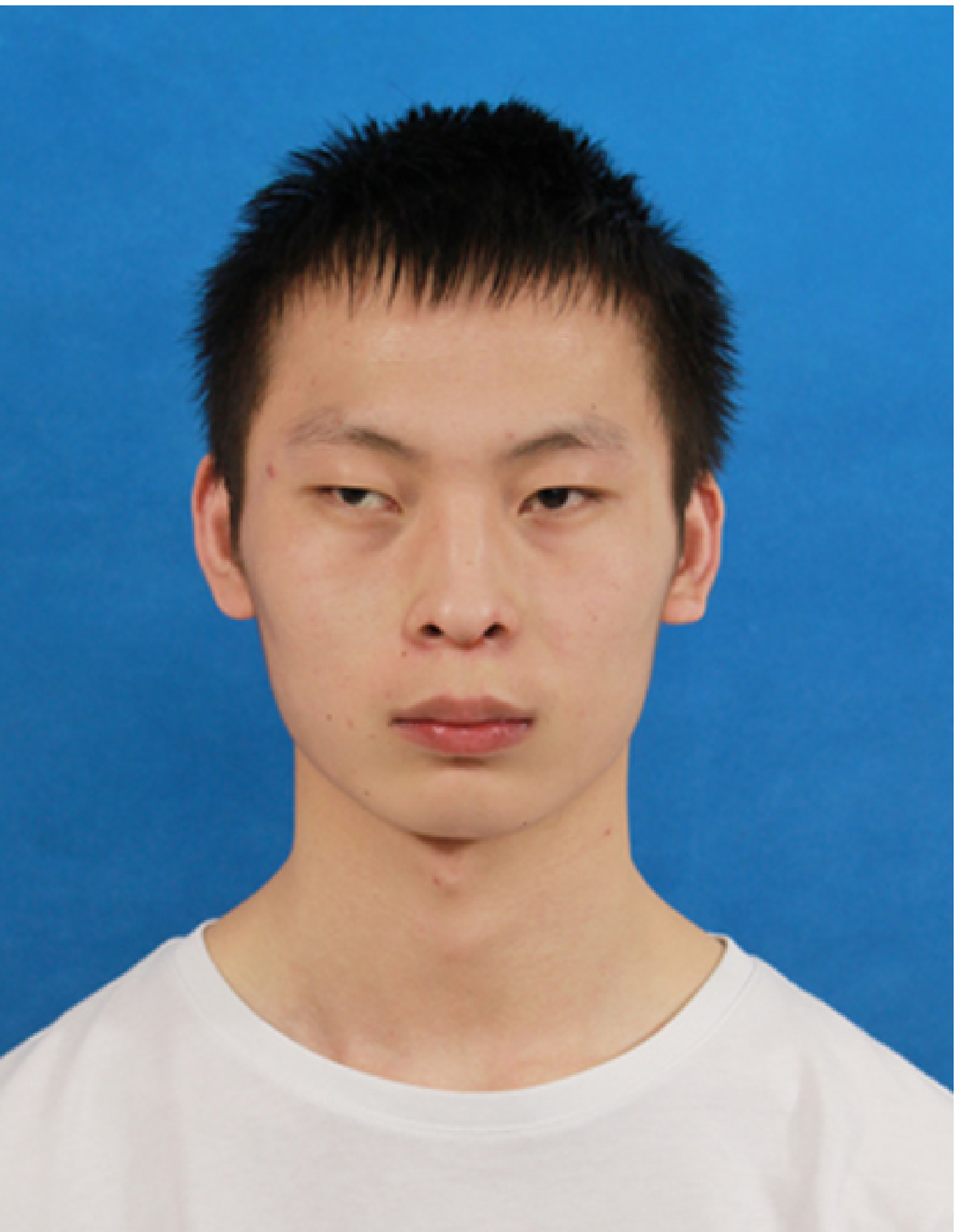}}]{Ren Ren} received the B.S. degree from the School of Mathematical Sciences, University of Electronic Science and Technology of China, Chengdu, in 2019. He is currently pursuing the M.S. degree in School of Automation Engineering, University of Electronic Science and Technology of China. His current research interests include network analysis and unsupervised learning.
\end{IEEEbiography}

\vspace*{-5ex}
\begin{IEEEbiography}[{\includegraphics[width=1in,height=1.25in,clip,keepaspectratio]{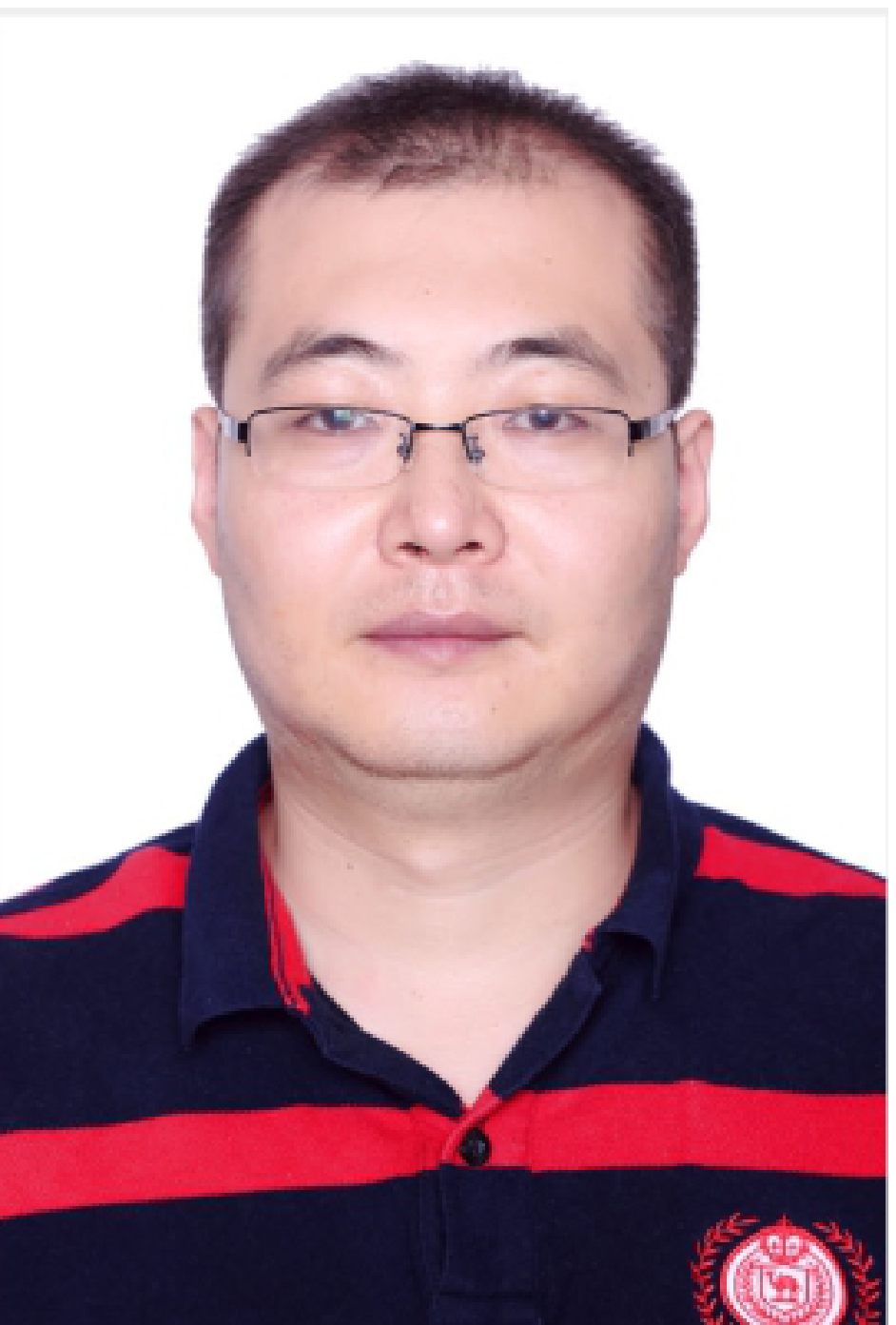}}]{Jinliang Shao} received the B.Sc. and Ph.D. degrees from the University of Electronic Science and Technology of China (UESTC), Chengdu, in 2003 and 2009, respectively. During 2014, he was a visiting scholar in Australian National University, Australia, and during 2018, he was a visiting scholar in Western Sydney University, Australia. He is currently 	an associate professor in the School of Automation	Engineering, UESTC. His research interests include 	multi-agent system, robust control, and matrix analysis with applications in control theory.
\end{IEEEbiography}

\vspace*{-5ex}
\begin{IEEEbiography}[{\includegraphics[width=1in,height=1.25in,clip,keepaspectratio]{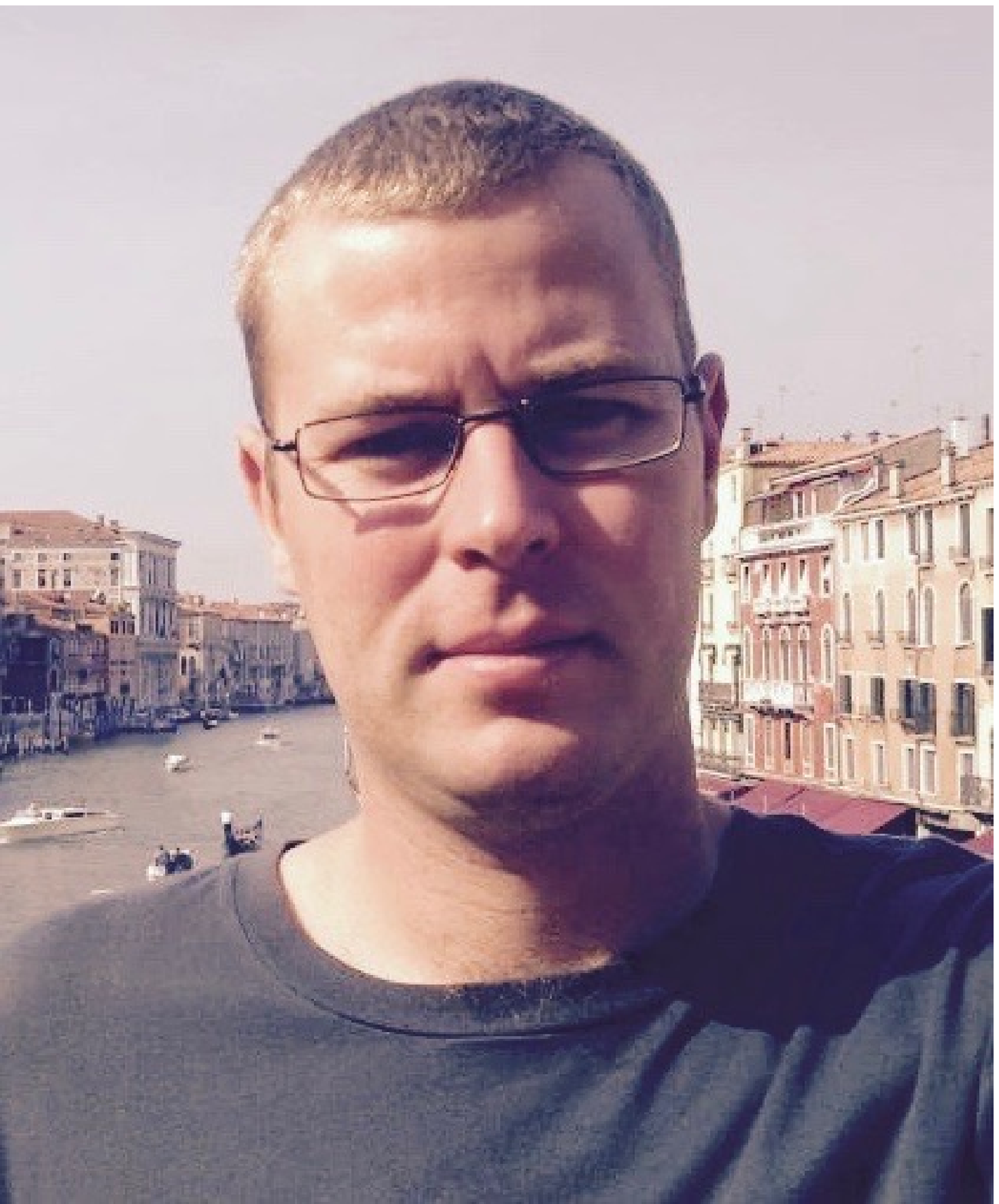}}]{Adrian N. Bishop}
	has held academic positions at the Royal Institute of Technology (KTH) in Stockholm, at the Australian National University (ANU) in Canberra and at the University of Technology Sydney (UTS) in Sydney. He is also a Research Scientist at NICTA in Canberra and Sydney. He is funded by an ARC Discovery Early Career Research Award (DECRA) Fellowship, NICTA and the US Air Force among other funding bodies. His current research interests fall within the intersection of statistical control and estimation, statistical machine learning and distributed (or large-scale) applicability of such topics.
\end{IEEEbiography}

\vspace*{-5ex}
\begin{IEEEbiography}[{\includegraphics[width=1in,height=1.25in,clip,keepaspectratio]{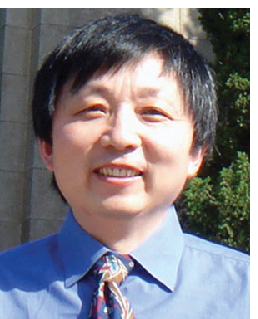}}]{Wei Xing Zheng} (M'93-SM'98-F'14) received the B.Sc. degree in Applied Mathematics, the M.Sc. degree and the Ph.D. degree in Electrical Engineering from Southeast University, Nanjing, China, in 1982, 1984, and 1989, respectively. He is currently a University Distinguished Professor with Western Sydney University, Sydney, Australia. Over the years he has also held various faculty/research/visiting positions at several universities in China, UK, Australia, Germany, USA, etc. He has been an Associate Editor of several flagship journals, including IEEE Transactions on Automatic Control, IEEE Transactions on Cybernetics, IEEE Transactions on Neural Networks and Learning Systems, IEEE Transactions on Control of Network Systems, IEEE Transactions on Circuits and Systems-I: Regular Papers and so on. He is a Fellow of IEEE.
\end{IEEEbiography}

\end{document}